\documentclass[manuscript, nonacm]{acmart}
\renewcommand\footnotetextcopyrightpermission[1]{} 
\usepackage{enumitem}
\usepackage{subcaption}
\newcommand{\xhdr}[1]{\vspace{1mm} \noindent{\bf #1}}


\usepackage{afterpage}

\setcitestyle{acmnumeric}

\newtheorem{finding}{Finding}
\newcommand\PaperTitle{Deconstructing the Filter Bubble: \\User Decision-Making and Recommender Systems}

\title[Deconstructing the Filter Bubble]{\PaperTitle}

\begin{document}


\begin{abstract}
We study a model of user decision-making in the context of recommender systems via numerical simulation.
Our model provides an explanation for the findings of Nguyen, et. al (2014), where, in environments where recommender systems are typically deployed, users consume increasingly similar items over time even without recommendation. We find that recommendation alleviates these natural filter-bubble effects, but that it also leads to an increase in homogeneity across users, resulting in a trade-off between homogenizing across-user consumption and diversifying within-user consumption. Finally, we discuss how our model highlights the importance of collecting data on user beliefs and their evolution over time both to design better recommendations and to further understand their impact.
\end{abstract}

\author{Guy Aridor}
\affiliation{%
  \institution{Columbia University}
  \city{New York}
  \state{NY}
}
\email{g.aridor@columbia.edu}

\author{Duarte Gon\c{c}alves}
\affiliation{%
  \institution{Columbia University}
  \city{New York}
  \state{NY}
}
\email{duarte.goncalves@columbia.edu}

\author{Shan Sikdar}
\affiliation{%
  \institution{Everquote}
  \city{Cambridge}
  \state{MA}
}
\email{shan.sikdar@gmail.com}

\maketitle

\section{Introduction}

Recommender systems (RS) have become critical for assisting users in navigating the large choice sets that they face on many online platforms. For instance, users have to choose from thousands of movies on Netflix, millions of products on Amazon, and billions of videos on YouTube. Users in many cases are not aware of most items, let alone have full information about their preferences over them. To make matters worse, the items in these contexts are usually experience goods whose true value for users can only be learned after consumption.
\par

RS have driven a significant fraction of consumer choice on these platforms with 75\% of movies watched on Netflix and 35\% of page-views on Amazon coming from recommendations.\footnote{MacKenzie et al. (2013, Oct.),  How retailers can keep up with consumers. \url{https://www.mckinsey.com/industries/retail/our-insights/how-retailers-can-keep-up-with-consumers}. Retrieved on October 3, 2019.} While there are many positive effects from these systems, there is an increasing worry about their unintended side-effects. There have been claims that personalized RS lead users into \textit{filter bubbles} where they effectively get isolated from a diversity of viewpoints or content \cite{pariser2011filter}, and that personalized RS may also lead users to become increasingly homogenized at the same time \cite{chaney2018algorithmic, hosanagar2013will}.
\par
Understanding how RS influence user behavior is important not only for characterizing the broader consequences of such systems but also for guiding their design. In this paper, we develop a theoretical model of user decision-making in contexts where RS are traditionally deployed. We utilize previous empirical studies that characterize how RS influence user choice as a benchmark and our theoretical model provides an intuitive mechanism that can explain these empirical results. The key insight of our model is that user \textit{beliefs} drive the consumption choices of users and that recommendations provide them with information that leads them to update their beliefs and alter their choices. A crucial component of our model is that users' beliefs about items are driven not only by recommendations, but also from their previous experiences with similar items. We use these insights to provide guidance for RS design, highlighting that understanding users' beliefs about the quality of the available items is essential to design recommendations and evaluate their impact.

\xhdr{Our Model.} We analyze a model of user choice with four central components.
\par
The first component of our model is that users sequentially consume items and face large choice sets. In our setting of interest, users are long-lived, but they only consume a small fraction of this choice set over their lifetime. This is traditionally the case on online platforms that have thousands or millions of options for users.
\par
The second component is that, prior to consuming them, users are \textit{uncertain} about how much they value the different items.
This is motivated both by the fact that recommender systems are traditionally deployed in contexts with experience goods, whose true value can only be learned after consumption, and the fact that such uncertainty is why RS exist in the first place. Thus, users face a sequential decision-making problem under uncertainty.
\par 
The third, and most crucial, element is that consumption of an item reveals information that changes user beliefs about their valuation of similar items. Unlike in standard sequential decision-making problems, once an item is consumed all uncertainty about its valuation is resolved and provides information that enables users to update their beliefs about similar items. This exploits the fact that the valuations of similar items are correlated which assists users in navigating the vast product space. The idea that users make similarity-based assessments to guide their choice has grounding in empirical evidence on how users navigate large choice sets \cite{schulz2019structured}.
\par 
Finally, in our model recommendation provides users with information about the true valuations. We model the realized valuations as being a weighted sum of a common-value and an idiosyncratic component. This formulation gives a stylized notion of predictability of user preferences where the idiosyncratic component is inherently unpredictable given other users' preferences and the common-value component is what the recommender can learn from previous users' data. We suppose that the recommender knows the common-value component for each item and combines it with users' beliefs over the product space when designing personalized recommendation.

\xhdr{Our Contributions.}
We provide a clear mechanism that explains the empirical results in \cite{nguyen2014exploring} who show that, in the context of movie consumption, user behavior is consistent with filter-bubble effects even without recommendation and that recommendation leads to users being less likely to fall into such filter bubbles. In this context, filter-bubble effects are defined as users consuming items in an increasingly narrow portion of the product space over time. The simple and intuitive driving force of this is that preferences for similar items are correlated, which implies that when an item is consumed and the user learns its value, it provides information about similar items. Crucially, this not only impacts the underlying belief about the expected value of similar items, but also how uncertain the user is about their valuation of them. Consequently, this learning spillover induces users to consume items similar to those they consumed before that had high realized value, leading to an increasing narrowing of consumption towards these regions of the product space. This effect is further amplified when users are \textit{risk-averse}, a concept from decision theory where all else being equal, users have a preference for items with lower uncertainty to those with higher uncertainty. However, by providing information to users, recommendation leads users to be more likely to explore other portions of the product space, limiting the filter bubble effect.
\par
We find that, while recommendation leads a single user to be more likely to explore diverse portions of the product space, it also coordinates consumption choices across users. This leads to an increase in homogeneity across users, resulting in a trade-off between homogenizing across-user consumption and diversifying within-user consumption. We explore the relationship between the overall diversity of consumed items and user welfare and find that more diverse sets of consumed items do not always correspond to higher user welfare. 
\par
Lastly, we discuss how our model and findings can be used to inform the design and evaluation of RS as well as the data that is traditionally collected for them. This highlights the importance of user beliefs in determining user consumption choices and how both recommendation and informational spillovers determine how these beliefs change over time. By collecting information on user beliefs, RS designers can understand what items a user would consume \textit{without} recommendation and then predict how providing information to the user would change her beliefs and resulting consumption decisions. Thus,  our evaluation measure determines the value of a recommendation based on the marginal welfare gain associated with providing a user with a recommendation over what the user would do without it. We discuss how this provides an additional rationale as to why ``accurate'' recommendations are not always good recommendations.
\section{Related Work.} 
\par
The first set of related works studies the extent and implications of filter bubbles. \cite{pariser2011filter} first informally described the idea of the ``filter bubble'' which is that online personalization services would lead users down paths of increasingly narrower content so that they would effectively be isolated from a diversity of viewpoints or content. Following this, a number of empirical studies in various disciplines, have since studied the extent to which this phenomenon exists in a wide range of contexts \cite{flaxman2016filter,hosanagar2013will,moller2018blame,nguyen2014exploring}. The most relevant to our study is \cite{nguyen2014exploring} who study whether this effect exists in the context of movie consumption. They find that even users whose consumption choices are not guided by recommendations exhibit behavior consistent with ``filter bubbles'' and that RS can actually increase the diversity of the content that users consume. To our knowledge there are no theoretical models that rationalize these empirical findings and we provide a theoretical framework through which to view this problem. Moreover, we provide a clear mechanism that drives such effects and how recommendation interacts with them.
\par 
Another set of papers has examined whether RS can lead users to become increasingly homogenized. \cite{celma2008hits, treviranus2009value} show that incorporating content popularity into RS can lead to increased user homogenization. \cite{chaney2018algorithmic} shows how user homogenization may arise from training RS on data from users exposed to algorithmic recommendations. \cite{fleder2009blockbuster} show that homogenization can increase due to a popularity recommendation bias that arises from lack of information about items with limited consumption histories. We show similar results as previous work where RS lead to increased user homogenization. However, the mechanisms behind this differ from existing work as homogenization arises due to the fact that recommendation leads users to coordinate their consumption decisions in certain portions of the product space.
\par
A third strand in the literature studies the impact of human decision-making on the design and evaluation of RS. \cite{chen2013human} surveys the literature on the relationship between human decision making and RS. The closest set of papers pointed out in this survey are those related to preference construction \cite{bettman1998constructive, lichtenstein2006construction} whereby users develop preferences over time through the context of a decision process. We point out that the true underlying preferences of users may be stable over time, but, due to the nature of items in contexts where RS are deployed, they have incomplete information of their valuations and both consumption and recommendation provide them with information to reduce their uncertainty. Thus, the primary insight of our paper is that user beliefs and how users update their beliefs about similar items after consumption are important and previously unconsidered elements of human decision making that are critical for understanding the design and consequences of RS. Within this literature, \cite{celma2008new, cremonesi2013user, pu2011user} focus on ``user-centric'' approaches to recommendation whereby user evaluation of the usefulness of recommendation is a key evaluation measure. Our evaluation measure is similar, but, unlike previous approaches, emphasizes the importance of user beliefs. Finally, \cite{hodgson2019horse} considers a similar model as ours where users engage in ``spatial learning'' and exploit the correlation of their preferences in the environment, but consider it in the context of search for a single item.

\section{Our Model and Preliminaries}
\subsection{Preliminaries on Expected Utility Theory}
\noindent For every item $n$ in the product space $\mathcal J$, we assume that each user $i$ assigns a monetary equivalent $x_{i,n} \in \mathbb R$ to the experience of consuming it. Each user can value the same item differently. However, we assume that users have the same utility over money, given by a utility function $u: \mathbb R \to \mathbb R$, strictly increasing and continuous. So, ex-post, the value of item $n$ for user $i$ is given by $u(x_{i,n})$. Before consuming the item, the user does not know exactly how she will value it. In particular, even users that will end up having the same ex-post valuation of item $n$ may differ in their ex-ante valuation because they hold different beliefs about it. We denote by $p_{i}$ the beliefs user $i$ has about how she will value each of the items in the product space. Note that this implies that consuming item $n$ is the same as taking a gamble. Each user evaluates the item according to the expected utility associated with the item, i.e. $U_i(n)=\mathbb E_{p_i}[u(x_n)]$. 
\par
Risk aversion captures how different users react to the risk associated to a particular consumption opportunity. It is formalized as follows: a given gamble $x$ takes real values and follows distribution $p$. Then, for every gamble $x$, there is a certain amount of money that makes the user indifferent between taking the gamble or taking the sure amount of money. This sure amount of money is called the \textit{certainty equivalent} of gamble $x$ and is denoted as $\delta(x)$. A user $i$ is more risk-averse than another user $j$ if whenever user $j$ prefers a sure thing to the gamble, then user $i$ does too. Therefore, a more risk-averse user is more willing to avoid the risk of taking the gamble. We assume that the utility function takes a flexible functional form $u(x)=1-\exp(-\gamma x)$ for $\gamma\ne0$ and $u(x)=x$ for $\gamma\to 0$ -- known as constant absolute risk-aversion preferences (from hereon CARA). Higher $\gamma$ implies higher risk-aversion, with $\gamma \to 0$ corresponding to the risk-neutral case and $\gamma>0$ to the risk-averse one. Our formulations here follow standard economic consumer theory (see \cite{mas1995microeconomic} for a textbook treatment of these topics).
\par

\subsection{Model}
\par
\xhdr{Users.} We consider a set of users $I$ where each user $i \in I$ faces the same finite set of $N$ items $\mathcal J = \left\{0,1,...,N-1\right\}$. For simplicity, we assume that users only derive pleasure from item $n \in \mathcal{J}$ the first time they consume it.
\par

We denote by $x_{i,n}$ user $i$'s realized value from consuming item $n$. In particular, we consider that the realized value derived from a given item can be decomposed in the following manner: $x_{i,n}= v_{i,n} + \beta v_n$, where $v_{i,n}$ denotes an idiosyncratic component -- i.e. user $i$'s idiosyncratic taste for item $n$ --  and $v_{n}$, a common-value component. One can interpret $v_n$ as a measure of how much item $n$ is valued in society in general and, in a sense, $v_{i,n}$ denotes how $i$ diverges from this overall ranking. The scalar $\beta \in \mathbb{R}_{+}$ denotes the degree to which valuations are idiosyncratic to each user or common across users. If $\beta=0$, it is impossible to generate meaningful predictions of any one's individual preferences based on others, while if $\beta$ is large, every individual has similar preferences.
\par
Stacking values in vector-form, we get the vector of values associated with each item 
$${\left(x_{i,n}\right)}_{n \in \mathcal{J}}=:X_i =V_i+ \beta V, $$
where $V_i ={\left(v_{i,n}\right)}_{n \in \mathcal{J}}$ and $V={\left(v_{n}\right)}_{n \in \mathcal{J}}$.
\par
\xhdr{User Decision-Making.}
We assume the user makes $T$ choices and therefore can only consume up to $T$ items, where $T$ is a small fraction of $N$. This captures the idea that users are faced with an immense choice set, but that ultimately they end up experiencing (and learning) about just a small fraction of it. For tractability, we impose that users are myopic and every period consume the item that they have not yet tried ($n_i^t$) that gives them the highest expected utility given the information from past consumption ($C_i^{t-1}=(n_i^1,...,n_i^{t-1})$) and their initial beliefs.
\par
\xhdr{User Beliefs.} We assume that all the true realized values are drawn at $t = 0$. However, users do not know the realized values before consuming an item, but rather have beliefs over them.
Formally, user $i$ starts with some beliefs about $X_i$, namely that the idiosyncratic and common-value parts of the valuations are independent -- $V_i \perp \!\!\! \perp V$ -- and that each is multivariate normal: 
\begin{enumerate}[topsep=0pt]
\item $V_i \sim \mathcal N (\overline V_i, \Sigma_i)$; and 
\item $V \sim \mathcal N(\overline V, \Sigma)$ with $\overline V =0$.
\end{enumerate}
We impose the normality assumption for two reasons. The first is that this allows for simple and tractable belief updating. The second is that it allows us to incorporate an easily interpretable correlation structure between the items. The precise formulation of $\Sigma$ and $\Sigma_i$ that we consider is defined below when we discuss user learning.
\par
Recalling that $V_i$ represents idiosyncratic deviations from $V$, we assume that, on the population level, prior beliefs $\overline V_i=\left(\overline v_{i,n}\right)_{n \in \mathcal{J}}$ are drawn independently from a jointly normal distribution, where $\overline v_{i,n} \sim \mathcal N (0, \overline \sigma^2)$ are independent and identically distributed. These $\overline v_{i,n}$ denote the prior belief that individual $i$ holds about her valuation over item $n$. As people are exposed to different backgrounds, their beliefs about how much they value a given item also varies and $\overline v_{i,n}$ denotes this idiosyncrasy at the level of prior beliefs.
\par

We assume users are expected utility maximizers. User $i$'s certainty equivalent for item $n$, the sure value that makes user $i$ indifferent between it and consuming the item $n$, conditional on the consumption history, is given by
$\delta_{i}(n)\mid C_i^{t-1}=\mu_n-\frac{1}{2}\gamma \Sigma_{nn}$, where $\mu_n$ and $\Sigma_{nn}$ are the expected value and variance for item $n$ that the user has given their initial beliefs and consumption history up until time $t$. Note that this expression is known to be the certainty equivalent for CARA preferences in a Gaussian environment \cite{mas1995microeconomic}. As it is immediate from this expression, the user assigns greater value to items for which the expected monetary equivalent, $\mu_n$, is higher, but penalizes those about which there is greater uncertainty $\Sigma_{nn}$, the more so the greater the user's degree of risk aversion $\gamma$.

\xhdr{User Learning.}
When a user consumes an item $n$ she learns the realized value for that item. We consider the case where  learning about the value of item $n$ reveals more about the value associated to items that are closer to it, which captures the idea that trying an item provides more information about similar items than about dissimilar ones.\footnote{\cite{schulz2019structured} empirically studies how individuals solve sequential decision-making problems under uncertainty in large choice sets in the context of mobile food delivery orders. They find that individuals engage in similarity-based generalizations where learning about the realized value of a particular item provides them with information about similar items. We incorporate this finding into our model in a stylized manner.} In order to have a well-defined notion of similarity we need to define a distance function between items, which we define as $d(n,m):=\min\{ \lvert m - n \rvert ,N - \lvert m - n \rvert \}$ where $m$ and $n$ are indices of items in $\mathcal{J}$. This distance function is not intended to model the intricacies of a realistic product space, but instead to provide a stylized product space to help us understand the effects of informational spillovers on user behavior. The basic intuition, in the context of movie consumption, is that a user's valuation of \textit{John Wick} (item $n$) provides more information about how much she will like \textit{John Wick: Chapter Two} (item $m$) than \textit{Titanic} (item $q$) since $d(n, m) < d(n, q)$. 
\par
We consider that the entry of $n$-th row and the ($m$)-th column of $\Sigma_i$ is given by $\sigma_i^2 \rho^{d(n,m)}$, and that of $\Sigma$ is given by $\sigma^2 \rho^{d(n,m)}$. The scalar $\rho \in [0,1]$ therefore impacts the covariance structure: a higher $\rho$ implies that learning the utility of $n$ is more informative about items nearby and, for $\rho \in (0,1)$, this effect is decreasing in distance. The particular distance function that we rely on leads to a simple covariance structure, where the $(n,n+1)$-th entry in the covariance matrix is $\sigma^{2} \rho$, the $(n,n+2)$-th entry is $\sigma^{2} \rho^2$, etc.\footnote{This exponential decay correlation structure can be related to the tenet of case-based similarity of \cite{gilboa1995case} -- see \cite{billot2008axiomatization} for an axiomatization of exponential similarity.}
\par
The precise updating rule is as follows. Recall that at time $t$ the user's consumption history is given by $C_{i}^{t}$ and we denote the utility realizations of these items as $c_t$. We denote $\mu_t$ as the initial mean beliefs the user has over the items in $C_{i}^{t}$ and $\mu_{N-t}$ as the initial mean beliefs the user has over the remaining $N-t$ items, $\mathcal{J} \setminus C_{i}^{t}$. We partition the covariance matrix as follows:
\[ \Sigma =  \left( \begin{array}{cc}
\Sigma_{(N-t, N-t)} & \Sigma_{(N-t,t)} \\
\Sigma_{(t,N-t)} & \Sigma_{(t,t)}
\end{array} \right).
\]
After consuming the items in $C_{i}^{t}$, the resulting beliefs over the remaining items are given by $\mathcal{N}(\bar{\mu}, \bar{\Sigma})$ where $\bar{\mu}$ and $\bar{\Sigma}$ are as follows:
\begin{align*}
\bar{\mu} \mid c_t &= \mu_{N-t} + \Sigma_{(N-t,t)} \Sigma_{(t,t)}^{-1}(c_t - \mu_t); \\
\bar{\Sigma} \mid c_t &= \Sigma_{(N-t,N-t)} - \Sigma_{(N-t,t)} \Sigma_{(t,t)}^{-1} \Sigma_{(t,N-t)}.
\end{align*}

\xhdr{An Illustrative Example.} We illustrate the main intuitions of our model with a simple example. Suppose that there are four items: 0, 1, 2, 3. The items are in different places of the product space, where 0 is close to 1 and 3 but more distant from 2. For the sake of expositional clarity, suppose that the initial mean beliefs are given by $\mu=(\mathbb E[x_n])_{n=0}^3=(0)_{n=0}^3$.
\par
In period 1, every item is ex-ante identical since they have the same mean and variance and so suppose that the user breaks the tie arbitrarily and consumes item 0. The underlying correlation structure implies that upon observing that $x_0 = y$ the user will update beliefs about the remaining three items according to the previously specified updating rule. For concreteness, we suppose that $\sigma = 1$ and $\rho = 0.5$, but the intuitions hold for any value of $\sigma$ and $\rho > 0$. First, we consider the case when the realization of $y > 0$ and, specifically, $y = 0.5$ -- though the general intuitions hold for any $y > 0$. The resulting beliefs after observing $y$ are then as follows:
\[ \bar{\mu} = (\mu \mid x_0=y) =  \left (\begin{array}{c}
\mathbb{E}[x_1\mid x_0=y] \vspace{0.15cm} \\
\mathbb{E}[x_2\mid x_0=y] \vspace{0.15cm} \\
\mathbb{E}[x_3\mid x_0=y]
\end{array}  \right) =\left (\begin{array}{c}
\rho y  \vspace{0.15cm} \\
\rho^{2} y  \vspace{0.15cm} \\
 \rho y \\
\end{array} \right) =
\left (\begin{array}{c}
\frac{1}{4} \vspace{0.15cm} \\
\frac{1}{8}  \vspace{0.15cm} \\
\frac{1}{4}
\end{array}  \right),\qquad 
\bar{\Sigma} =(\Sigma\mid x_0=y)=  \left( \begin{array}{ccc}
\frac{3}{4} & \frac{3}{8} & 0 \vspace{0.15cm} \\
\frac{3}{8} & \frac{15}{16} & \frac{3}{8} \vspace{0.15cm}  \\
0 &\frac{3}{8} & \frac{3}{4}  \\
\end{array} \right).
\]
Thus, upon learning $x_0=y$, the user updates beliefs about the remaining items. Note that $\mathbb{E}[x_1\mid x_0=y] = \mathbb{E}[x_3\mid x_0=y] > \mathbb{E}[x_2\mid x_0=y]$ since item 0's value is more informative about similar items' values, items 1 and 3, than items further away in the product space such as item 2. Moreover, $\bar{\Sigma}_{11} = \bar{\Sigma}_{33} < \bar{\Sigma}_{22}$ as the uncertainty about items 1 and 3 is further reduced compared to item 2. Thus, since $y > 0$, the user in the next period will consume items nearby to item 0 since, even though initially she believed that all items had the same mean, the spillover from consuming item 0 leads her to believe that items 1 and 3 have higher expected valuations. Since both the mean is higher for these items and the variance is lower, the user will consume items 1 and 3 regardless of her risk aversion level.
\par 
Now we consider the case when item 0 ends up having a negative valuation so that $y = -0.5 < 0$. This results in $\mathbb{E}[x_1\mid x_0=y] = \mathbb{E}[x_3\mid x_0=y] = -\frac{1}{4} <  -\frac{1}{8} = \mathbb{E}[x_2\mid x_0=y]$ with $\bar{\Sigma}$ remaining the same as when $y = 0.5$. In this case the risk-aversion levels of the user determine the choice in the next period. If the user is risk-neutral ($\gamma = 0$), then she will go across the product space to consume item $2$ in the next period since it has a higher expected value. However, if she is sufficiently risk-averse then she may still consume item $1$ or $3$ since her uncertainty about these items is lower than item $2$. In particular, this will happen when 
\begin{align*}
\delta(3) = \delta(1) = \rho y - \frac{1}{2} \gamma \bar{\Sigma}_{11} > \rho^{2} y - \frac{1}{2} \gamma \bar{\Sigma}_{22} = \delta(2)
\end{align*}
Given the aforementioned parametrization and $y = -0.5$, the user will consume item $1$ or $3$ when $\gamma > \frac{4}{3}$ and will consume item $2$ when $\gamma < \frac{4}{3}$. Thus if the user is risk averse enough, then she might be willing to trade-off ex-ante lower expected values for lower risk and stick to consuming nearby items just because these items have lower uncertainty. 
\par 
This example illustrates the main mechanisms that can lead to excessive consumption of similar items. Once the user finds items in the product space with high valuations she will update her beliefs positively about items in this portion of the product space and continue consuming these items regardless of her level of risk aversion. However, this same updating leads to a reduction in uncertainty of these items and so, if she is sufficiently risk-averse, she still may continue consuming items in this portion of the product space, even if she has bad experiences with them, since they are perceived to be less risky. 
\par

\xhdr{Recommendation.}
Our model of recommendation is stylized in order to provide qualitative insights into how recommendation shapes behavior, instead of focusing on the details of how RS are implemented in practice. We model recommendation as giving users information about the valuation of the items.
\par

We will consider three cases. The case of primary interest is \textit{recommendation} where the recommender observes values accrued and knows $V$ but does not know $V_i$.\footnote{We do not consider the acquisition of information for the recommender to know $V$ and suppose that she has sufficient data to learn $V$ with arbitrary precision at $t = 0$.} However, the recommender does know the users' beliefs $\bar V_i$. Thus, at any given period, the recommender provides a personalized recommendation that combines the knowledge of the common value component $V$ with the user beliefs $\bar V_i$. Knowing the user's beliefs about her valuation of each item become crucial in this case: just providing the user with information about $V$ may change the user's original ranking, but, without considering the user's beliefs, she will not necessarily follow the recommendation.\footnote{The notion of recommendation that we consider is idealized where the recommendation does the Bayesian updating for users, but the results are equivalent to if the users did the updating themselves.}
\par

We further consider two cases that serve mainly as benchmarks. The first is \textit{no recommendation}, where users get no additional information and make consumption choices based on their beliefs and consumption history. This gives us a benchmark as to how users would behave \textit{without} recommendation so that we can analyze what changes with the introduction of recommendation. The second is the \textit{oracle recommendation} where the recommender knows the true realized utility of each item for each user and can therefore recommend the best remaining item in every period. This gives us a full information benchmark, which is the optimal consumption path for a user if all uncertainty about their preferences was resolved. Comparison to the oracle regime benchmark provides an analog to the standard regret measures utilized in the multi-armed bandit literature, which look at the difference in the expected value of the ex-post optimal action and the expected value of actions that were taken.
\par

\xhdr{Simulation Details.}
We analyze our model using numerical simulation since the sequential decision-making component paired with the rich covariance structure between the items make it difficult to characterize optimal user behavior analytically.\footnote{The Gaussian assumption allows for closed form belief updating which enables us to simulate our model but does not provide much help in analytical characterizations.}
\par

We explore how consumption patterns differ as we consider different recommendation regimes and report representative results from our simulations. We run this simulation over 100 populations of users with 100 users per population for each combination of parameters. 
A given set of parameters and a user are a single data point in our dataset.
\par

We simulate over risk-aversion parameters $\gamma \in \{ 0, 0.3, 0.6, 1, 5 \}$, \ standard-deviation $\sigma \in \{ 0.25, 0.5, 1.0, 2.0, 4.0 \}$, correlation parameters $\rho\in \{ 0, 0.1, 0.3, 0.5, 0.7, 0.9 \} $ and degree of idiosyncrasy of individual valuations $\beta \in \{ 0, 0.4, 0.8, 1, 2, 5\}$. The range of considered parameter values cover the relevant portions of the parameter space in order to provide full insight into the behavior of the model. When we consider results varying a single parameter, we group the results over the other parameters and provide reports varying only the parameter of interest. We report results for a relatively small consumption history $T=20$ with a product space size $N=200$.\footnote{Due to space constraints, we only report the values for $N = 200$. However, we further conducted the same exercises for $N = 100$ and $N = 500$ and the results are qualitatively similar to those reported here.}
\section{Results}
\subsection{Local Consumption and Filter Bubbles}
We characterize ``filter bubble'' effects as the degree to which users engage in \textit{local consumption}. We define local consumption in terms of the average consumption distance between the items consumed by the users at time $t-1$ and $t$. Thus, in the context of our model, filter-bubble effects arise when the average consumption distance decreases over time and, across regimes, when the levels are lower for a given recommendation regime compared to another.

Our first finding can be summarized as follows:

\begin{finding}\label{finding_local_consumption}
The impact of recommendation on local consumption:
\begin{enumerate}
\item When $\rho = 0$, there is no difference in consumption distance between the three recommendation regimes.
\item When $\rho > 0$, no recommendation induces more local consumption than both recommendation and oracle regimes. This effect is amplified as $\rho$ increases as well as when users are more risk averse ($\gamma$ increases).
\end{enumerate}
\end{finding}
\afterpage{
\begin{figure}[t]
\caption{Local Consumption and Correlation}
\begin{subfigure}{.3\linewidth}
  \centering
  \includegraphics[width=1.0\linewidth]{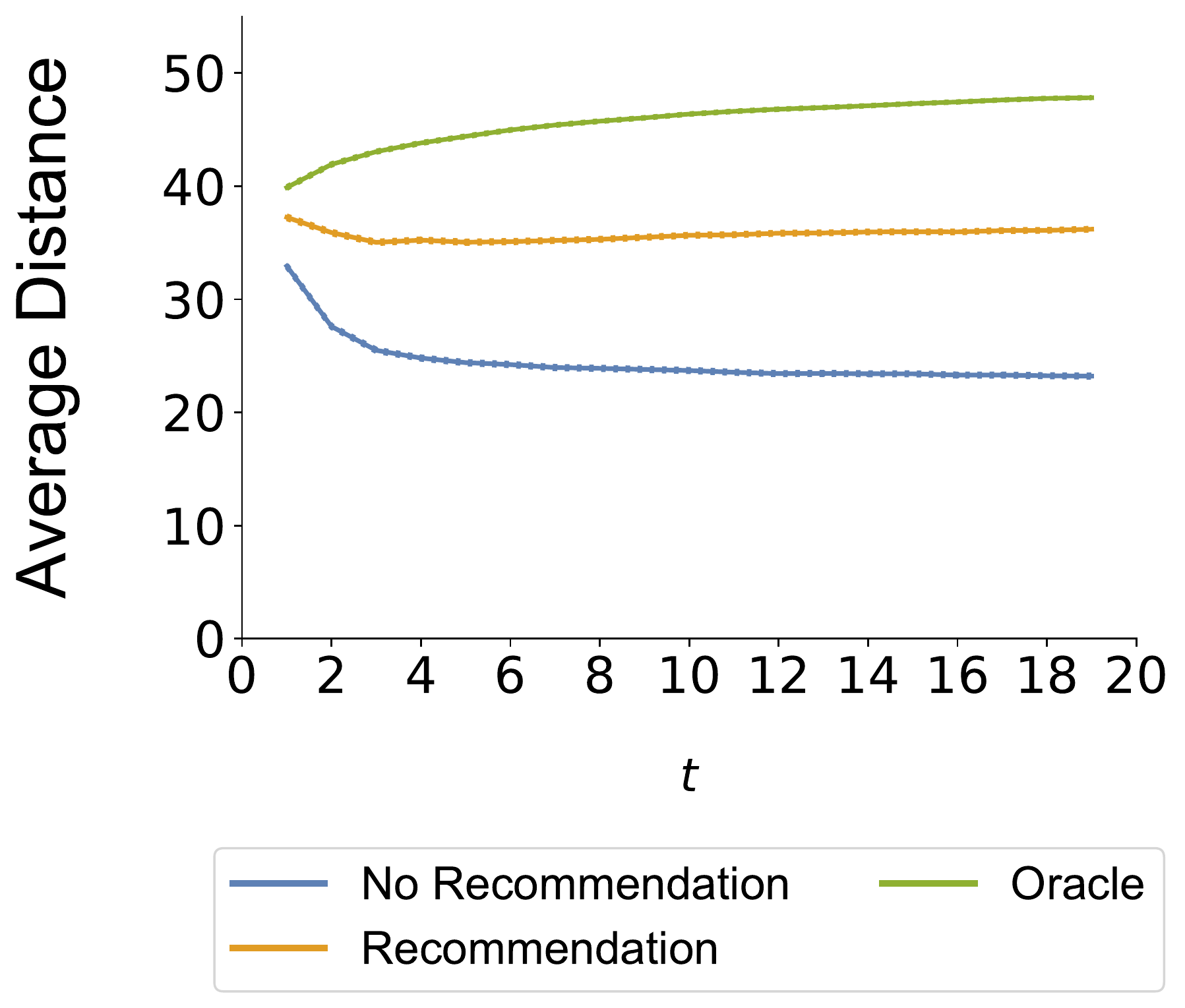}
  \label{fig:sfig1}
\end{subfigure}%
\hspace{1cm}
\begin{subfigure}{.3\linewidth}
  \centering
  \includegraphics[width=1.0\linewidth]{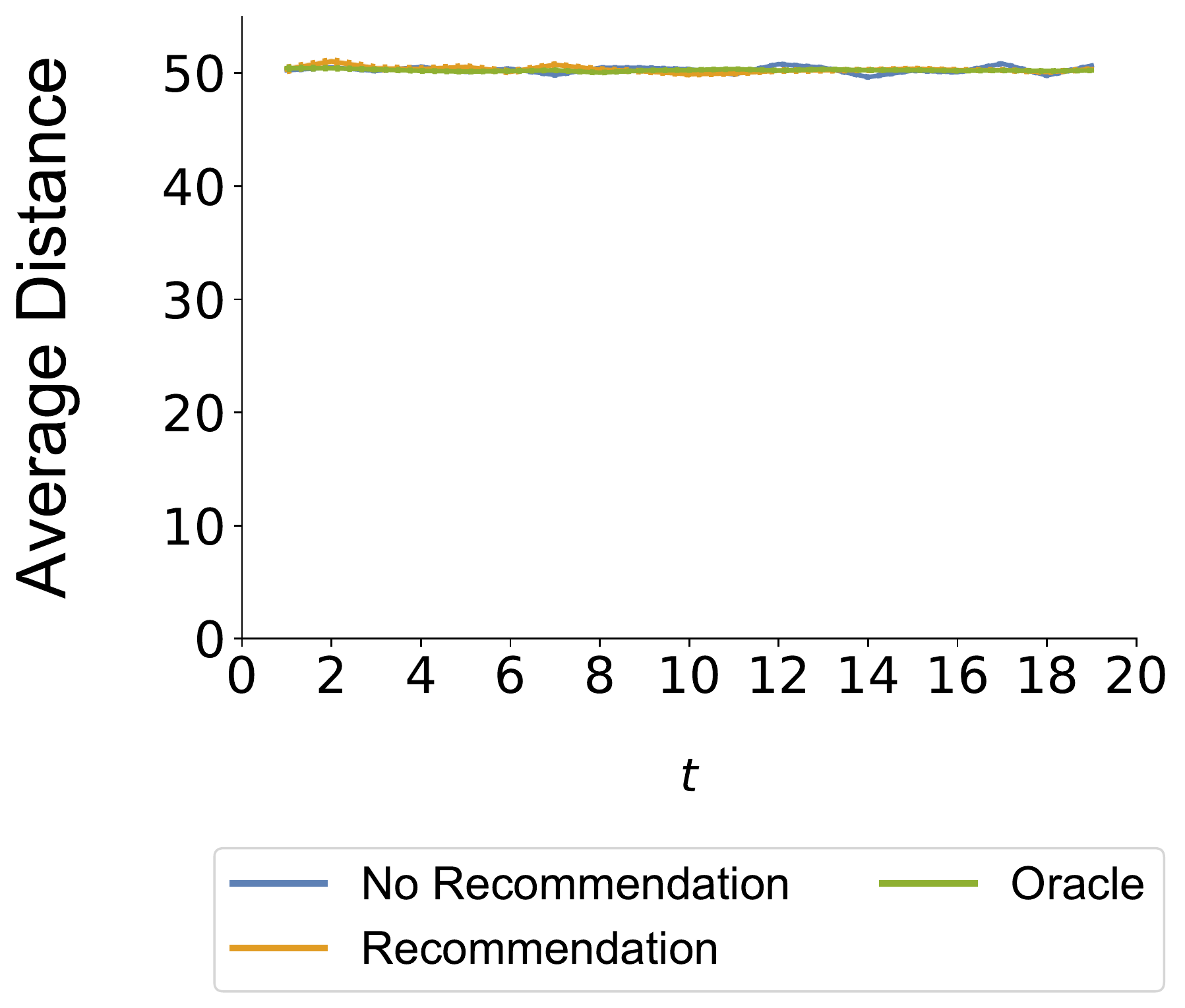}
  \label{fig:sfig2}
\end{subfigure}
\caption*{\scriptsize Notes: The figure shows the consecutive consumption path difference between the no recommendation, recommendation, and oracle regime. The figure on the left displays the average consecutive consumption distance aggregating over simulations where $\rho \in (0,1)$ and the figure on the right displays the average consecutive consumption distance aggregating over simulations where $\rho = 0$. The shaded area represents the 95\% confidence interval.}
\label{fig:correlation_consumption_path}
\end{figure}
\addtocounter{figure}{-1}}

First, the right panel of Figure \ref{fig:correlation_consumption_path} shows that, when $\rho = 0$, there is no difference in consumption distance between the three regimes. This is due to the fact that when $\rho = 0$, there is no reason that items that are close in the product space should have similar values and so the optimal consumption path does not depend on the similarity of the items. However this also means that users do not learn anything about neighboring items and so there is limited path-dependence in consumption. Not only is there no difference in the levels between the three regimes, but they all have the same, flat, average consumption distance path. This underscores the fact that if there were no correlation between the realized utilities then there would be no reason for users to consume similar items and thus no narrowing effect, regardless of the information on the true utility of the items that users had.
\par
The left panel of Figure~\ref{fig:correlation_consumption_path} shows that, when $\rho \in (0,1)$, both recommendation and no recommendation lead to increasingly local consumption compared to the oracle benchmark case. Moreover, the average consumption path between periods is \textit{decreasing} for the no recommendation case whereas it is \textit{increasing} for the oracle case. The recommendation regime decreases the degree of local consumption, but not as much as the oracle benchmark. Due to the correlation of values, the oracle consumption path exploits this and leads to the consumption of more similar items than in the case when $\rho = 0$. However, since these spillovers also impact user learning in the no recommendation case, users \textit{over-exploit} these and increasingly consume items similar to high value items that they have consumed before. This is also illustrated in the top row of~\autoref{fig:local_consumption_vary_rho_gamma}, which shows how the consumption paths in the oracle and no-recommendation regimes vary as $\rho$ increases and is in line with this intuition.
\par
As shown in the bottom row of~\autoref{fig:local_consumption_vary_rho_gamma}, this effect is further amplified as the level of risk aversion increases: the degree of local consumption drastically increases as $\gamma$ increases. This is due to the fact that the spillovers not only impact the mean expected belief about quality but also the degree of uncertainty. Local consumption therefore leads to users having less uncertainty about certain areas of the product space and risk aversion may lead them to increasingly consume nearby items.

In sum, filter-bubble effects only arise when there is an inherent correlation between the realized utilities of the items in the product space. When there is a correlation between the realized utilities then filter bubbles can naturally arise due to the nature of how individuals acquire additional information about the remaining items. We have shown that, unless users are provided with additional information to guide their consumption choices, then these information spillovers and user risk-aversion can lead users into filter bubbles. When users consume high valued items, they exploit the underlying correlation across different items' values, stronger for similar items, which leads them to increasingly consume items in narrower and narrower portions of the product space. Risk aversion may lead users into performing local consumption even when they have a low valuation of nearby items just because they know what to expect from the item. Recommendation leads to these effects being mitigated by providing users with additional information on items outside the already explored portions of the product space. Logically, if all uncertainty were resolved as in the oracle regime, then such behavior is not present.

\subsection{User Welfare and Item Diversity}

In this section we primarily focus on the impact of recommendation on user welfare and the overall diversity of the items that they consume. While in the previous section we looked at the distance between consecutive items, in this section we focus on a diversity measure that considers the entire consumed set of items. The diversity measure we utilize is common in RS literature (e.g. \cite{ziegler2005improving}) which is the average normalized pairwise distance between the consumed items:
$$D_i:=\frac{1}{N}\frac{1}{T(T-1)}\sum_{n,m \in C_i^T: n \ne m} d(n,m)$$
\noindent 
Finding~\ref{finding_diversity} summarizes the main results on item diversity:
\begin{finding}\label{finding_diversity}
The impact of recommendation on item diversity:
\begin{enumerate}
\item When $\rho = 0$, item diversity is the same across all three recommendation regimes;
\item When $\rho \in (0,1)$, item diversity decreases across all recommendation regimes but decreases the most in the no-recommendation regime. This effect is amplified as $\rho$ increases as well as when users become more risk-averse.
\end{enumerate}
\end{finding}
\par 

\begin{figure}[t]
\centering
\caption{Relationship between Local Consumption and Correlation ($\rho$), Risk Aversion ($\gamma$)}
\begin{subfigure}{.3\linewidth}
\includegraphics[width=\linewidth]{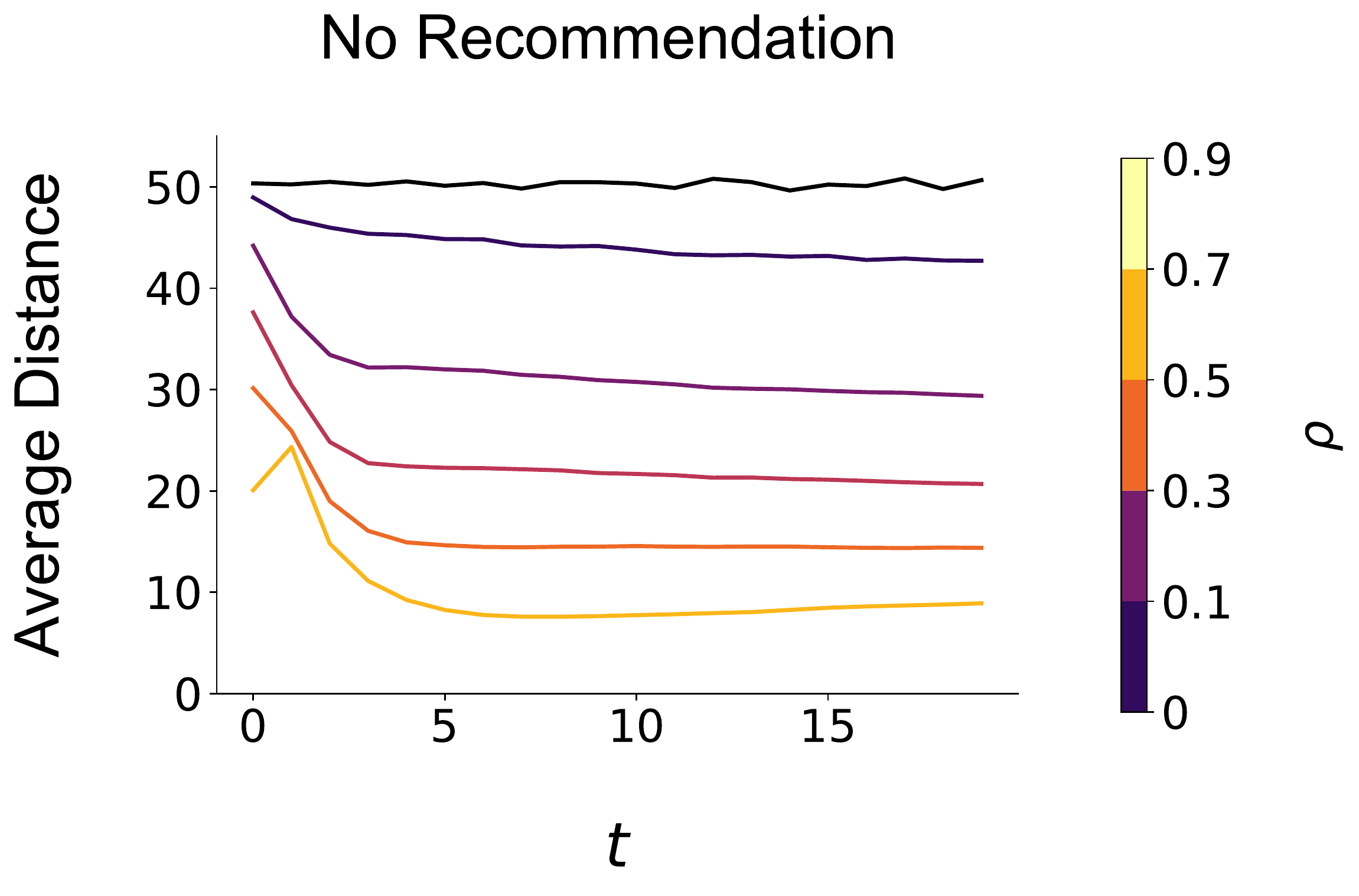}
\end{subfigure}
\begin{subfigure}{.3\linewidth}
\includegraphics[width=\linewidth]{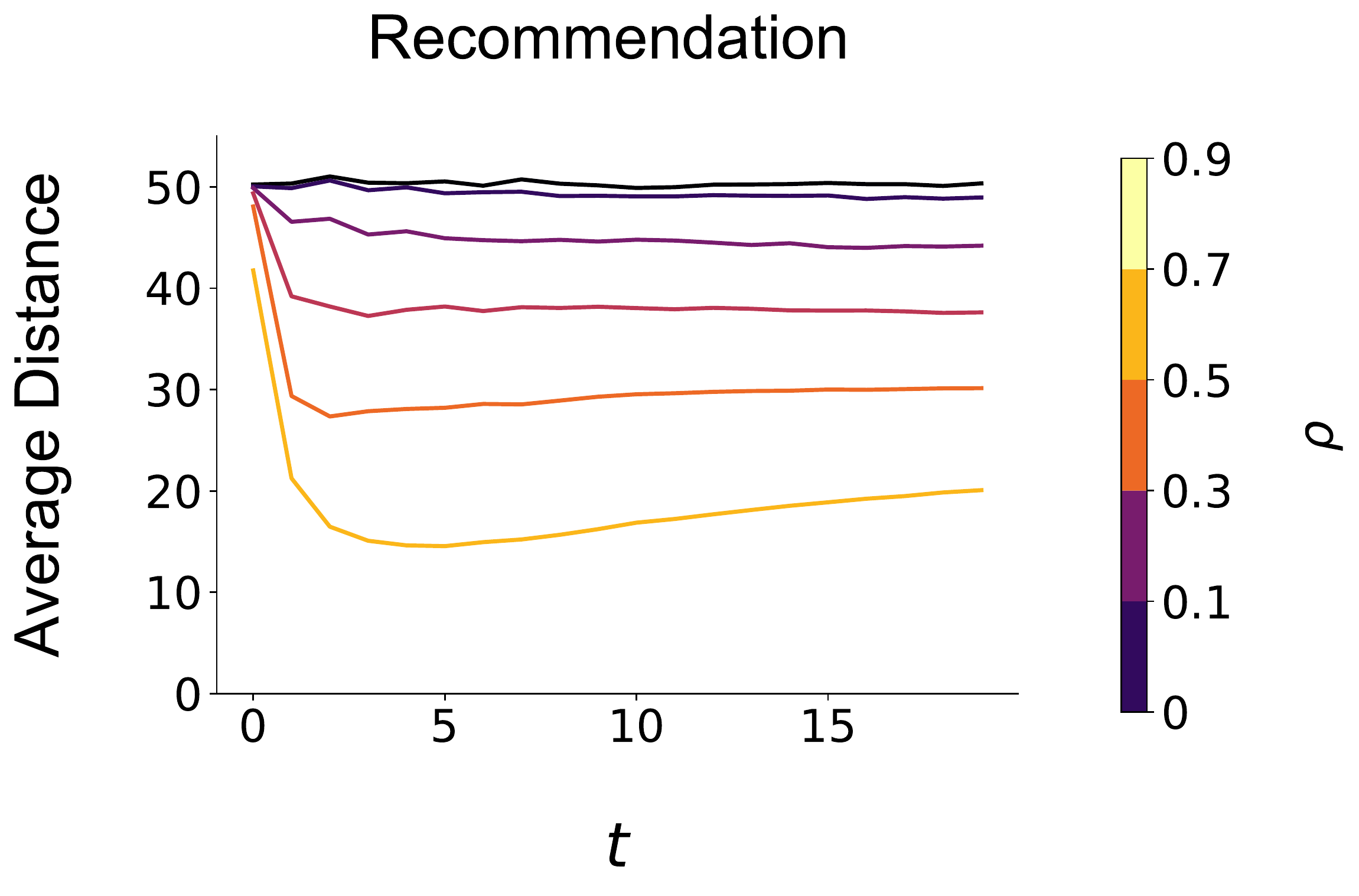}
\end{subfigure}
\begin{subfigure}{.3\linewidth}
\includegraphics[width=\linewidth]{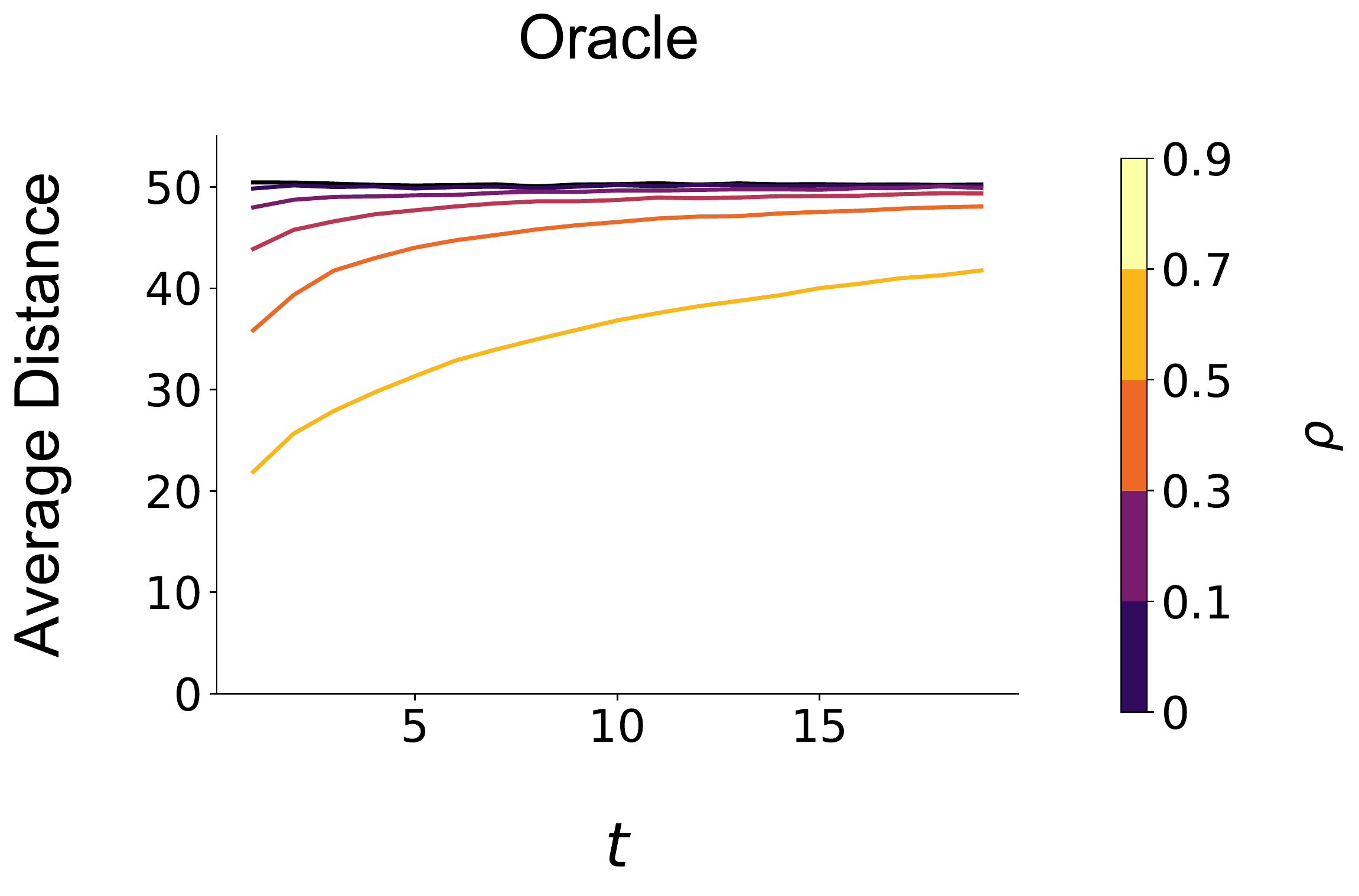}
\end{subfigure}\\
\begin{subfigure}{.3\linewidth}
\includegraphics[width=\linewidth]{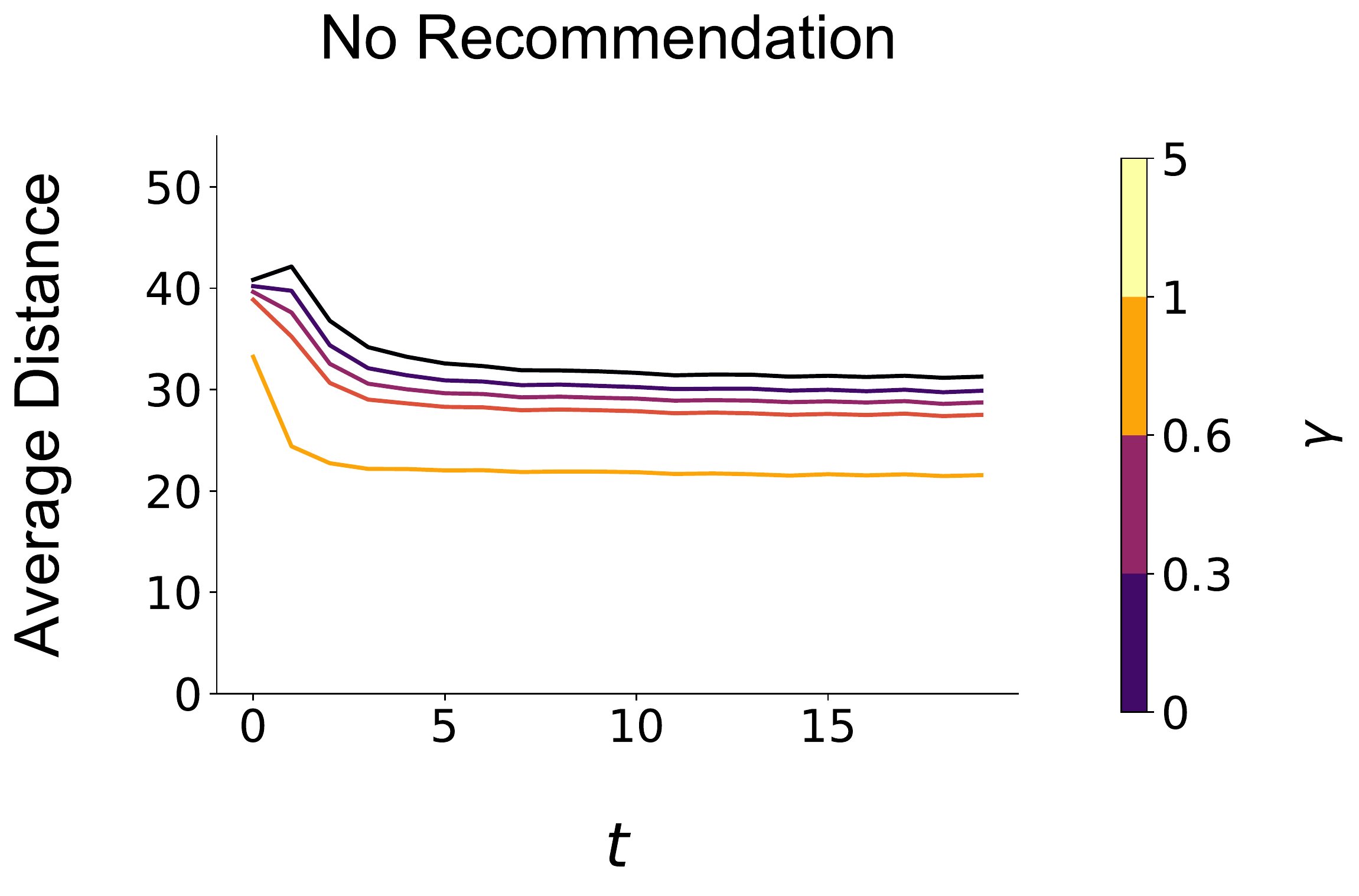}
\end{subfigure}
\begin{subfigure}{.3\linewidth}
\includegraphics[width=\linewidth]{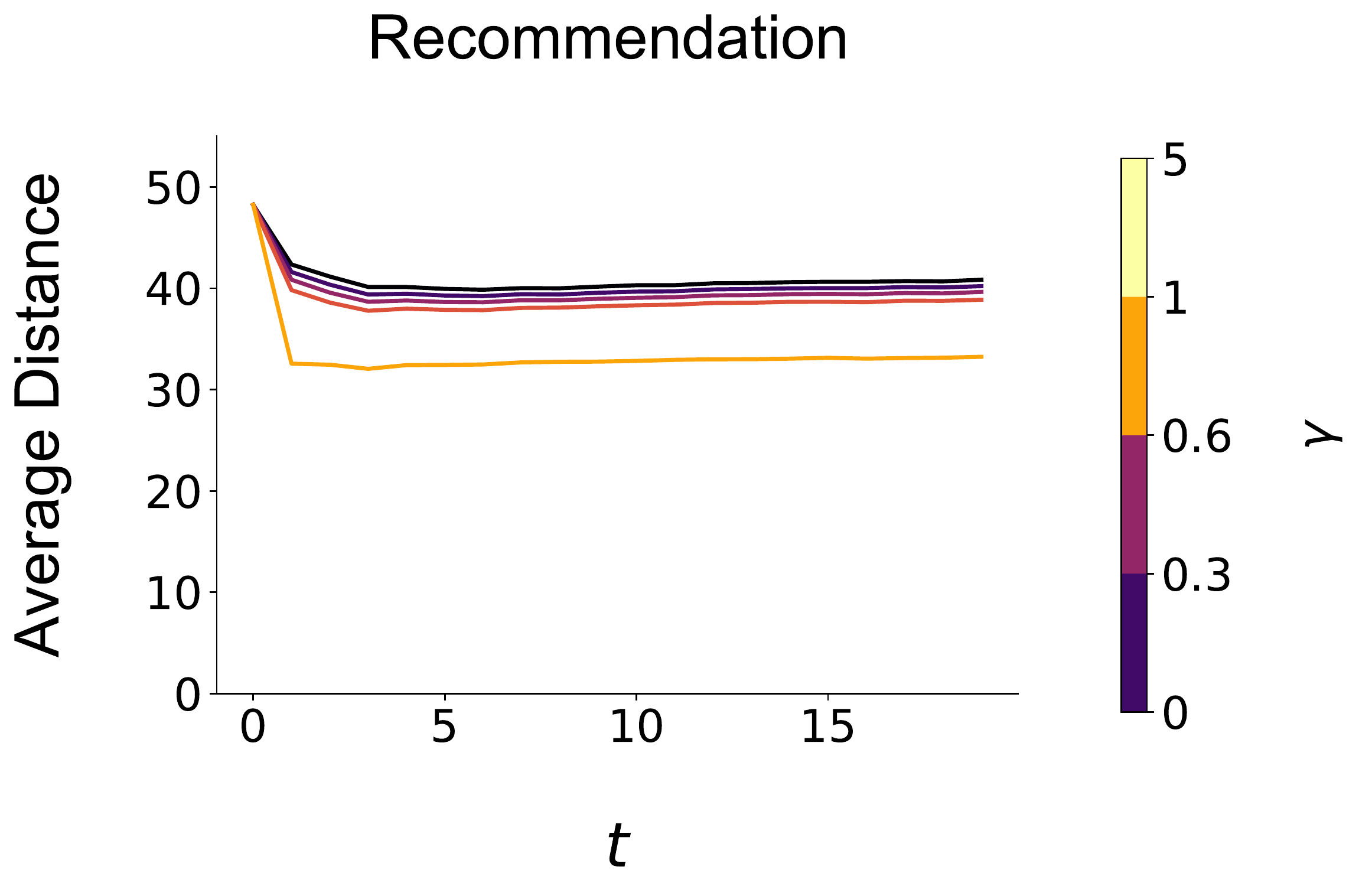}
\end{subfigure}
\begin{subfigure}{.3\linewidth}
\includegraphics[width=\linewidth]{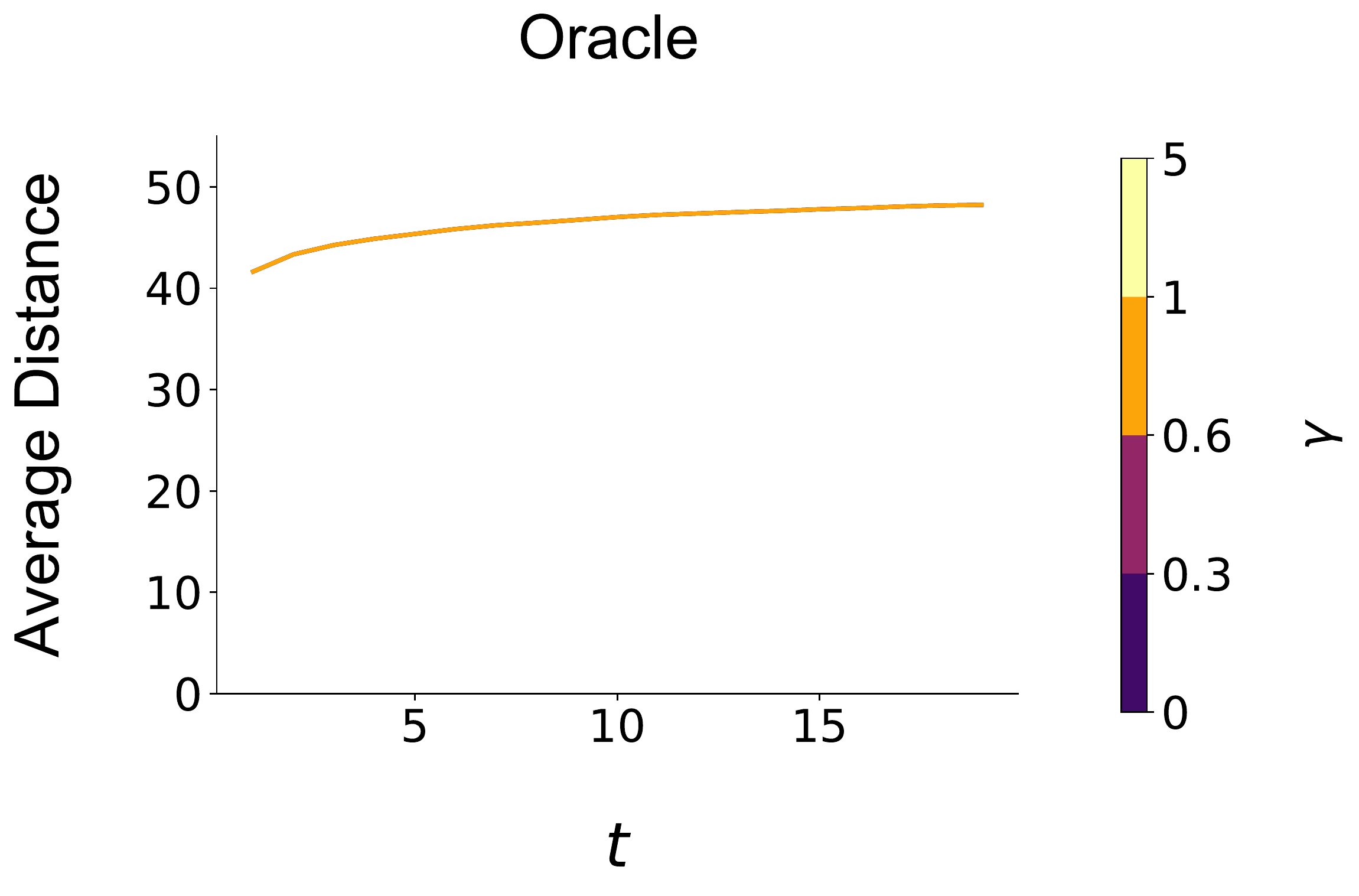}
\end{subfigure}%
\caption*{\scriptsize Notes: Each figure plots the average consecutive consumption distance across time as the inherent correlation between the valuation of the items, $\rho$, varies (top row) and the level of risk-aversion, $\gamma$, varies (bottom row). In both rows, the left displays the no-recommendation regime, the center displays the recommendation regime, and the right displays the oracle regime.}
\label{fig:local_consumption_vary_rho_gamma}
\end{figure}
\addtocounter{figure}{-1}

As before, when there is no correlation between valuations, item diversity is the same across different recommendation regimes. The over-exploitation of information spillovers when $\rho \in (0,1)$ leads to item diversity being lowest in the no-recommendation regime. As a result, this effect gets amplified as $\rho$ increases, which leads to the gap in diversity between the regimes to increase as $\rho$ increases. The top row of~\autoref{fig:local_consumption_vary_rho_gamma} shows how diversity varies as $\rho$ increases across the three regimes that we consider. There is a similar increasing diversity gap as $\gamma$, or the level of risk-aversion, increases as can be seen in the bottom row of~\autoref{fig:local_consumption_vary_rho_gamma}. The mechanisms behind these effects directly parallels those discussed in the previous section since low average sequential consumption distance directly leads to low diversity.
\par 

We now study how recommendation impacts user welfare. In our model users make consumption decisions that maximize their current period ex-ante utility that depends on their beliefs in that period. Thus, from an ex-ante perspective, they make optimal decisions, but our primary interest is in understanding how the ex-post, or realized, utility varies across regimes and parameter values. We define user's \textit{ex-post} welfare as the average of the realized values, controlling for the effect of $T$:
$$W_i:= \frac{1}{T}\sum_{n \in C_i^T} x_{i,n}$$

\begin{figure}[t]
\caption{Relationship between User Welfare, Diversity and Correlation ($\rho$), Risk Aversion ($\gamma$)}\label{fig:diversity_welfare_correlation}
\begin{subfigure}{0.3\linewidth}
  \includegraphics[width=1.0\linewidth]{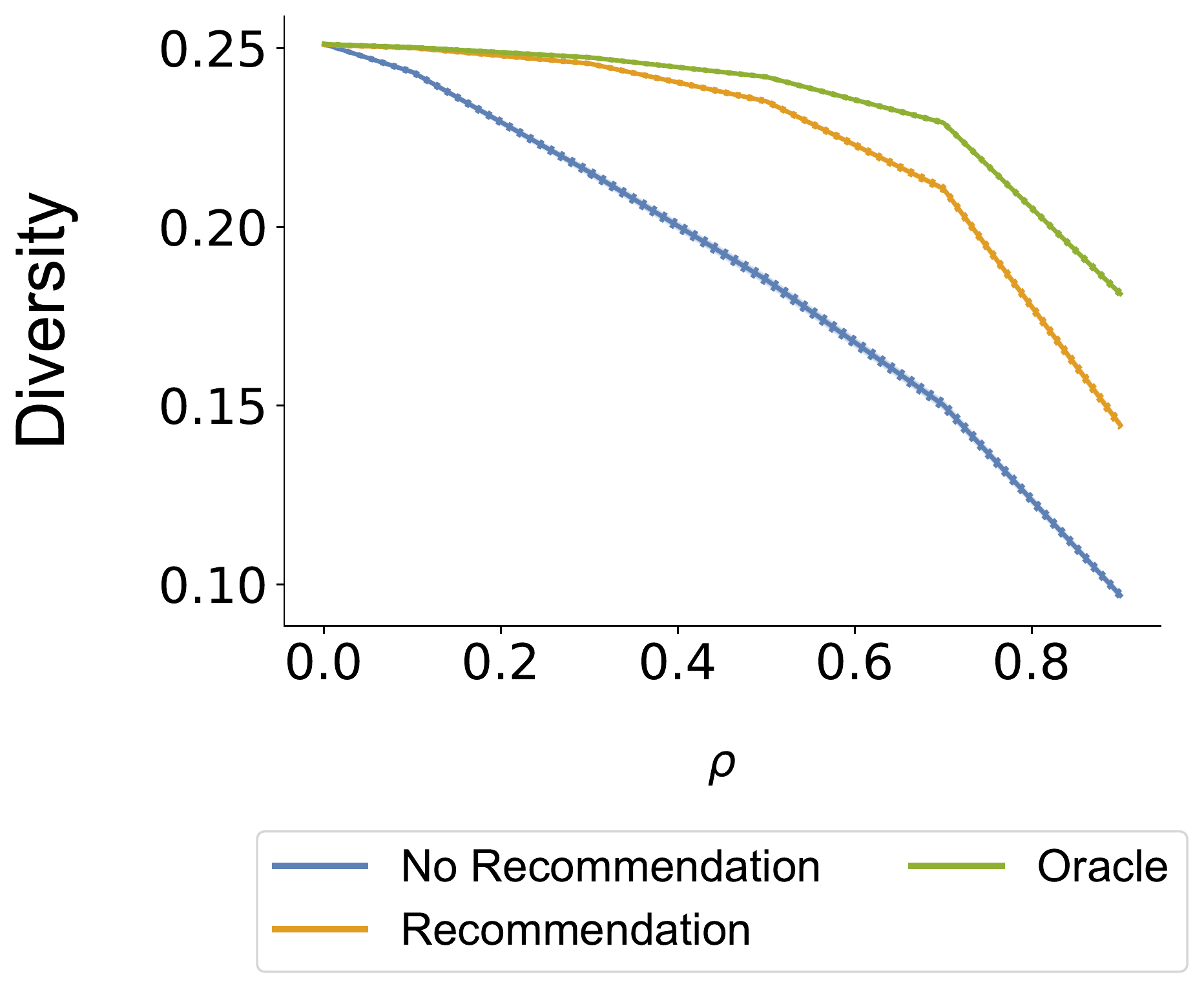}
\end{subfigure}
\begin{subfigure}{0.3\linewidth}
\includegraphics[width=1.0\linewidth]{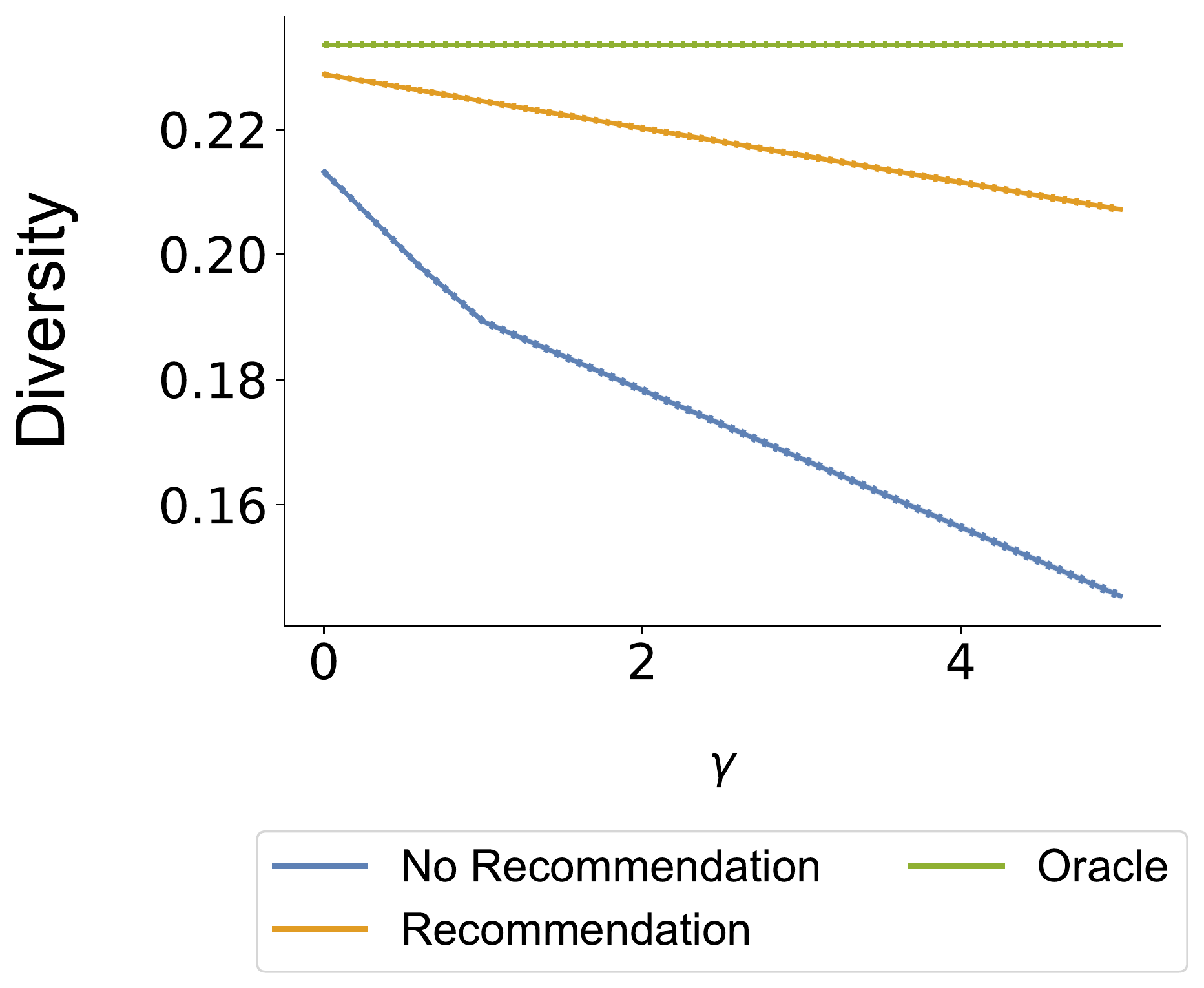}
\end{subfigure}
\begin{subfigure}{0.3\linewidth}
  \includegraphics[width=1.0\linewidth]{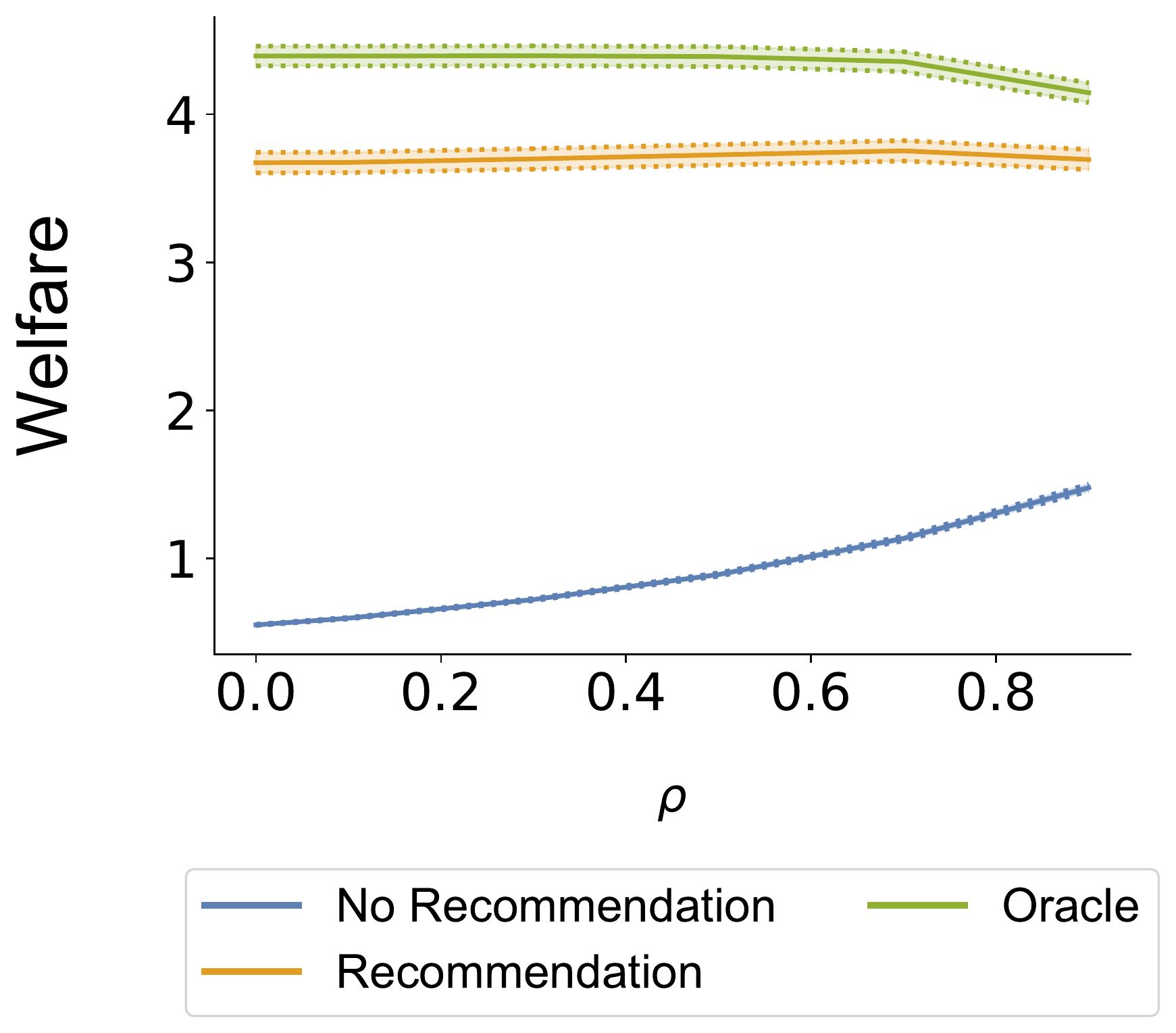}
\end{subfigure}
\caption*{\scriptsize Notes: The figures on the left and center display the relationship between $\rho$ and overall consumed item diversity (left) as well as $\gamma$ and overall consumed item diversity (center). The figure on the right displays the relationship between $\rho$ and overall welfare. The shaded area represents the 95\% confidence interval.}
\end{figure}

\noindent Finding~\ref{finding_welfare_gap} states our main findings of the impact of recommendation on ex-post welfare, which can be seen in the rightmost plot of~\autoref{fig:diversity_welfare_correlation}:

~\par
\begin{finding}\label{finding_welfare_gap}
The impact of recommendation on consumer welfare is as follows:
\begin{enumerate}
\item Under oracle recommendation, welfare is invariant with respect to $\rho$.
\item Under no recommendation, welfare is increasing in $\rho$.
\item Recommendation introduces welfare gains relative to no recommendation, but these gains are decreasing as $\rho$ increases.
\end{enumerate}
\end{finding}
\par 

\addtocounter{figure}{-1}

The most interesting observation is that the value of recommendation decreases as $\rho$ decreases. While welfare in the recommendation regime is flat as we increase $\rho$, it is increasing in the no-recommendation regime and thus the welfare gap between the two shrinks as $\rho$ increases. The intuition is clear as recommendation provides users with information that allows them to better guide their decisions and increase welfare. However, as $\rho$ increases users get increasingly more information from consuming items since the realized utility is now more informative about the utility of nearby items. Thus, since consumption decisions themselves yield valuable information, the information provided by recommendation is less valuable to the user.
\par 
One striking observation is that the decrease in diversity does not appear to be associated with a decline in welfare. Indeed, it appears that the opposite is the case - that low diversity is associated with higher welfare and vice versa. We next explore the relationship between welfare and diversity.
\par
\noindent Finding~\ref{finding_diversity_welfare_corr} summarizes our findings on the relationship between diversity and welfare:

\begin{finding}\label{finding_diversity_welfare_corr}
In the no-recommendation regime, diversity and welfare are:
\begin{enumerate}
\item Negatively correlated when users have no risk-aversion;
\item Uncorrelated when users have high levels of risk-aversion.
\end{enumerate}
In the recommendation regime, diversity and welfare are:
\begin{enumerate}
\item Uncorrelated when users have no risk-aversion;
\item Positively correlated when users have high levels of risk-aversion.
\end{enumerate}
In the oracle regime, diversity and welfare are always uncorrelated.
\end{finding}

\begin{figure}[t]
\caption{Diversity vs. Welfare}
\begin{minipage}{1.2\textwidth}
\hspace*{-1cm}\begin{subfigure}{.23\textwidth}
\includegraphics[width=1.0\linewidth]{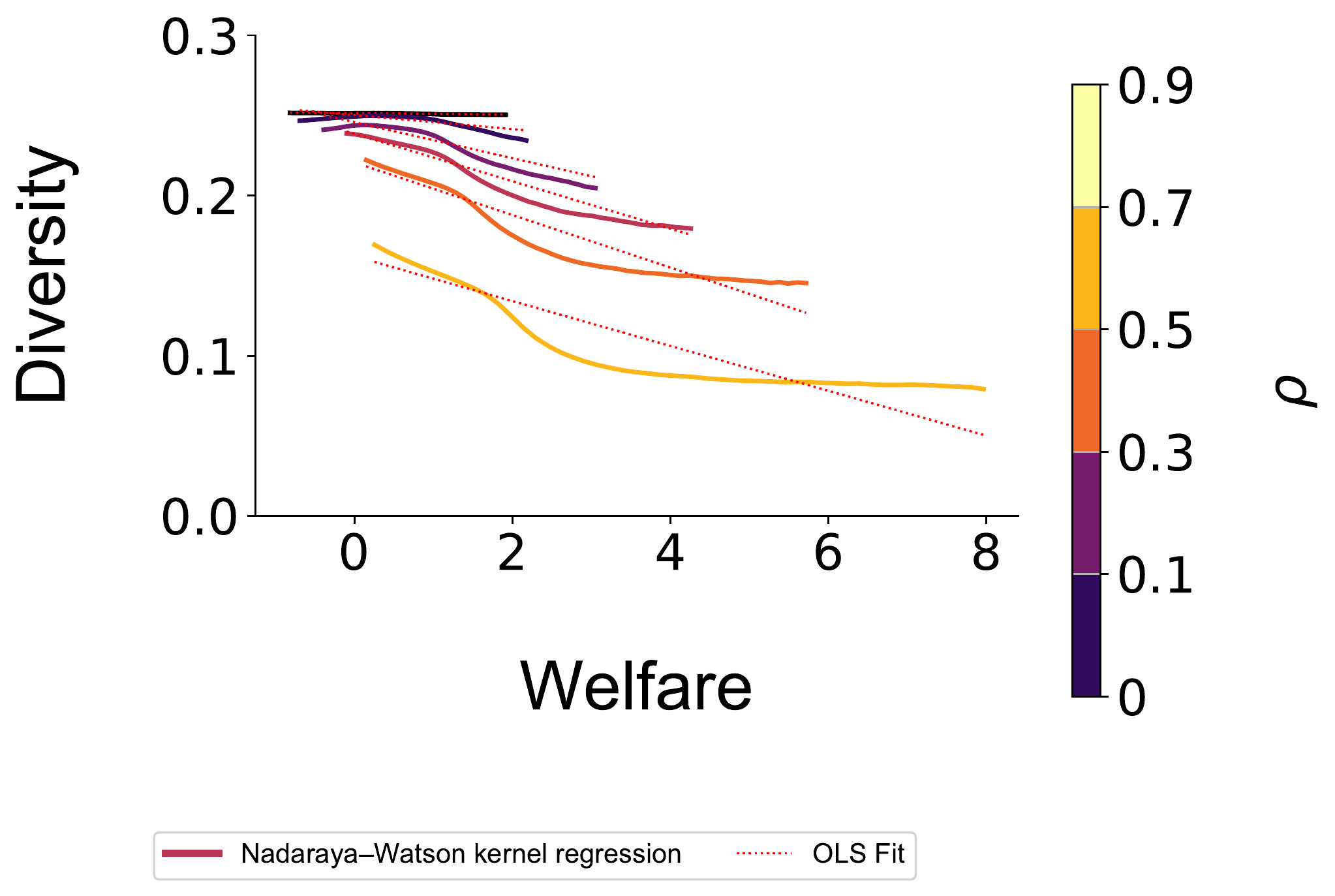}
\end{subfigure}
\begin{subfigure}{.23\textwidth}
\includegraphics[width=1.0\linewidth]{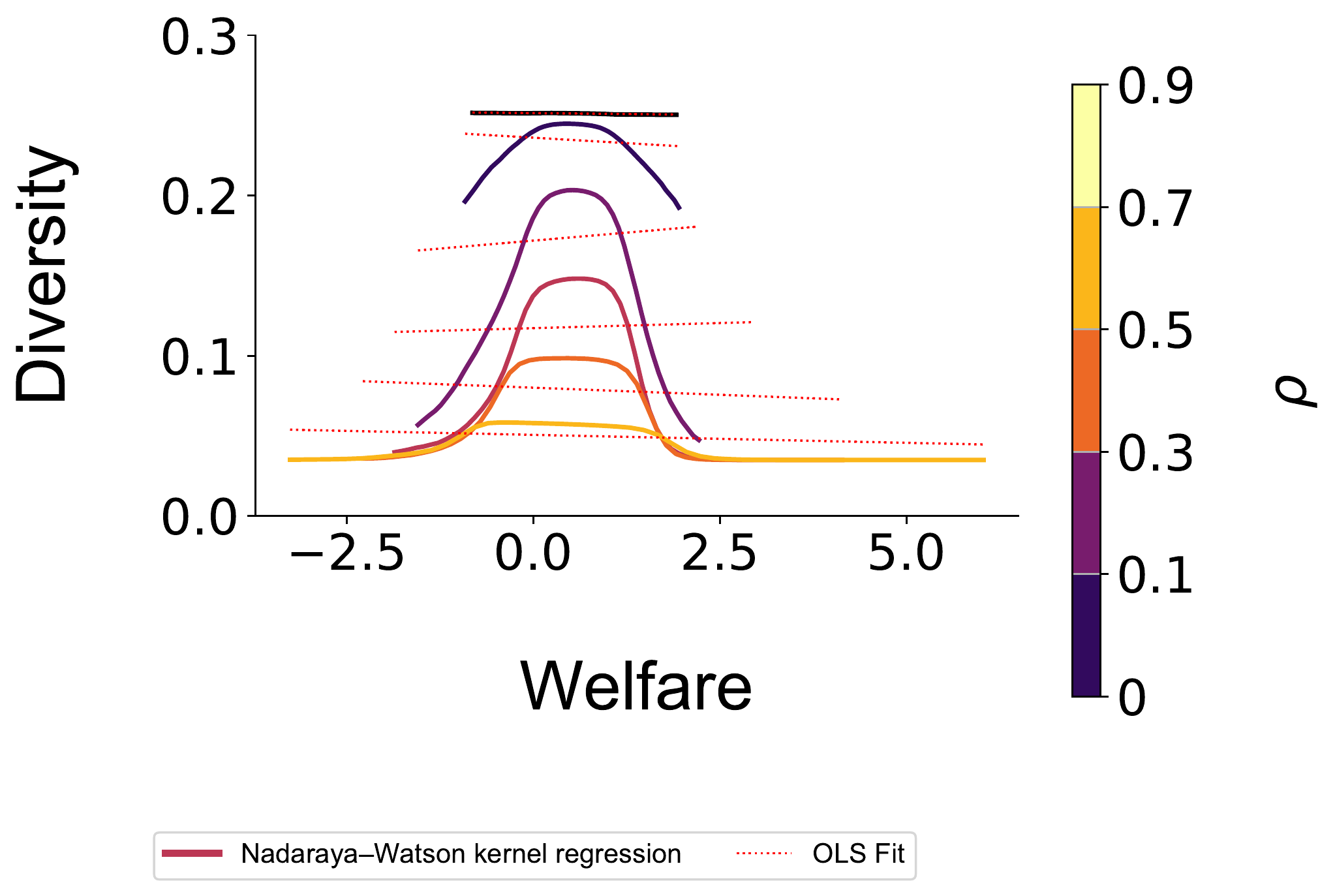}
\end{subfigure}
\begin{subfigure}{.23\textwidth}
\includegraphics[width=1.0\linewidth]{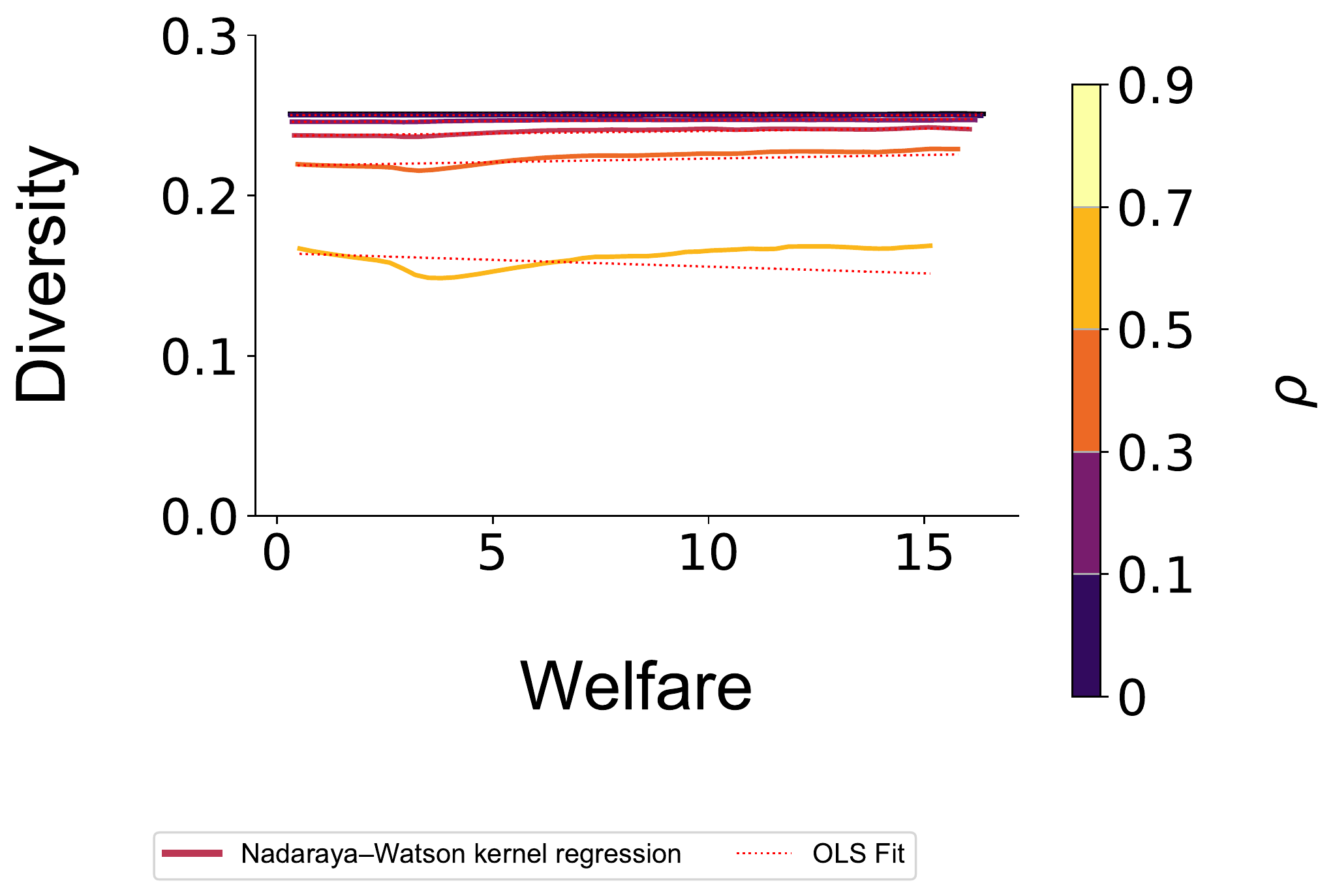}
\end{subfigure}
\begin{subfigure}{.23\textwidth}
\includegraphics[width=1.0\linewidth]{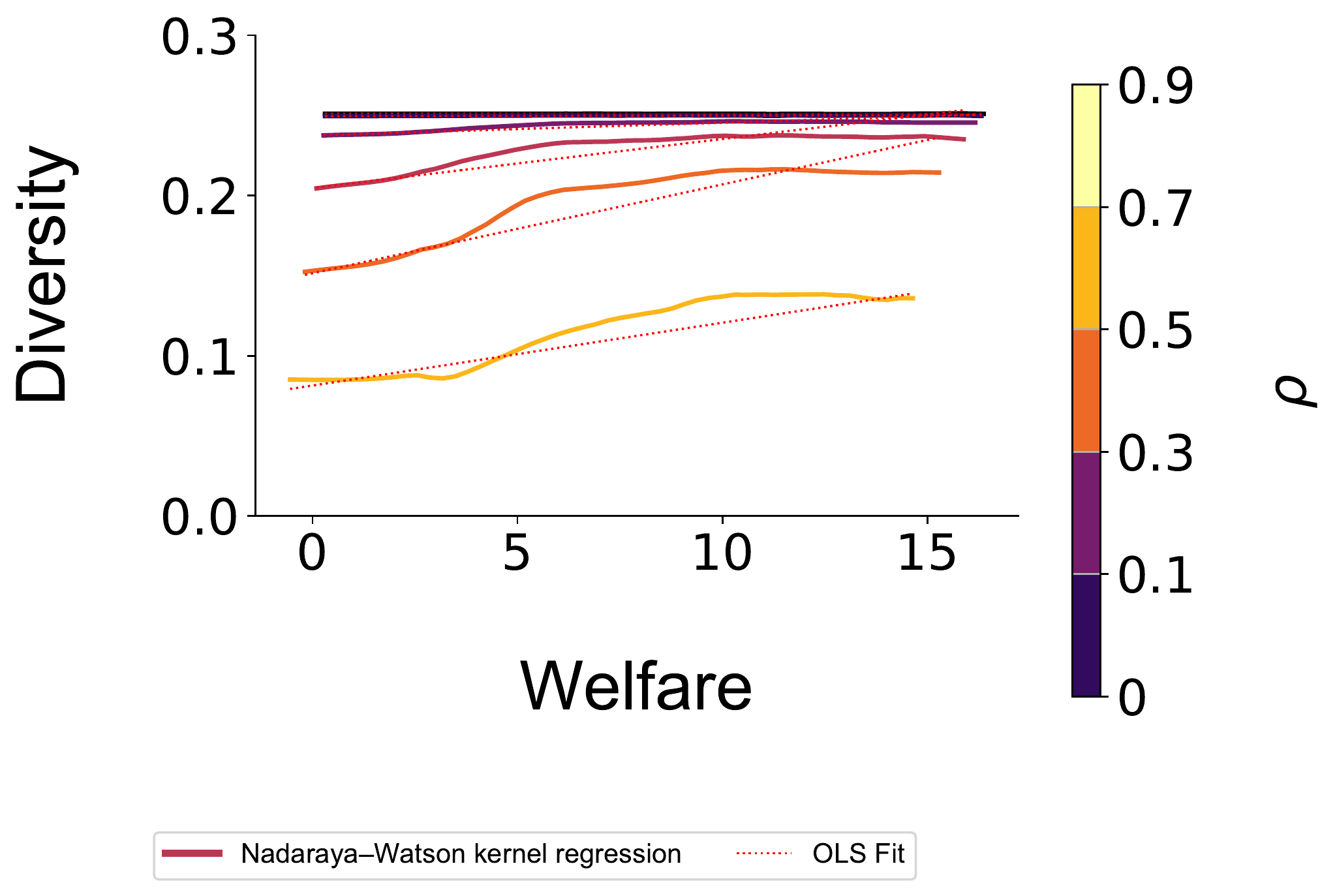}
\end{subfigure}
\end{minipage}\\
\caption*{\scriptsize Notes: The figure plots the relationship between diversity and welfare under no recommendation, with $\gamma = 0$ (first) and $\gamma = 5$ (second), and under recommendation, with $\gamma = 0$ (third) and $\gamma = 5$ (fourth).}\label{fig:diversity_welfare_ra}
\end{figure}
\addtocounter{figure}{-1}

Figure \ref{fig:diversity_welfare_ra} shows how diversity and welfare correlate for the no recommendation case as we vary the degree of risk aversion. When there is no risk-aversion then there is a negative correlation between welfare and diversity. This is since, with no risk-aversion, a user will select the item that she currently believes has the highest expected value. High item diversity in this case can arise from a user who consumes an item she disliked and updates her beliefs about nearby items negatively. As a result, in the following period she will pick an item far away in the product space from the item that was previously consumed. If instead the user valued highly the item that she had consumed, then she is more likely to pick a nearby item. The information spillovers therefore lead to high item diversity being negatively correlated with welfare.
\par
This only happens since $\gamma = 0$ leads to users only caring about the expected value of the item. However, as we saw in Findings \ref{finding_local_consumption} and \ref{finding_diversity}, increasing $\gamma$ can lead to lower diversity and increasingly local consumption due to the fact that the degree of uncertainty now impacts users' choices. This weakens the negative relationship between diversity and welfare since both negative and positive experiences with an item reduce uncertainty about surrounding items. This leads to the inverted-U shape found in Figure \ref{fig:diversity_welfare_ra} when $\gamma$ is relatively large (e.g. $\gamma = 5$) though diversity and welfare are virtually uncorrelated in the data. In the recommendation and oracle regimes, under risk neutrality ($\gamma=0$), welfare and diversity are uncorrelated, while under risk aversion ($\gamma=5$), it is possible to observe an actual positive relation between diversity and welfare as recommendations are able to reduce uncertainty and facilitate exploration of the product space.

\subsection{User Homogenization}

In this section, we focus on comparisons across users and investigate how the consumed set of items across users varies across different recommendation regimes and parameter values. In particular we look at the degree of \textit{homogenization} between users. Similar to other papers that study the degree of homogenization in RS (e.g. \cite{chaney2018algorithmic}) we measure homogeneity via the Jaccard index between the consumption sets of users:
\begin{align*}
H:=\frac{1}{|I|(|I|-1)}\sum_{i,j \in I: i \ne j}d_J(C_i^T,C_j^T)
\end{align*}
where $d_J$ denotes the Jaccard index and $H \in [0,1]$.
\par
\noindent Finding~\ref{finding_homogeneity} summarizes our findings on the impact of recommendation on user homogeneity:

\begin{finding}\label{finding_homogeneity}
The impact of recommendation on homogeneity is as follows:
\begin{enumerate}
\item Highest under recommendation and lowest under no recommendation;
\item Increasing in $\beta$, or the weight of the common-value component;
\item Decreasing in $\rho$ for partial recommendation, but weakly increasing in $\rho$ for no recommendation.
\end{enumerate}
\end{finding}

\begin{figure}[t]
\caption{Relationship between Homogeneity and Common-Value Strength ($\beta$), Correlation ($\rho$)}
\includegraphics[width=.3\linewidth]{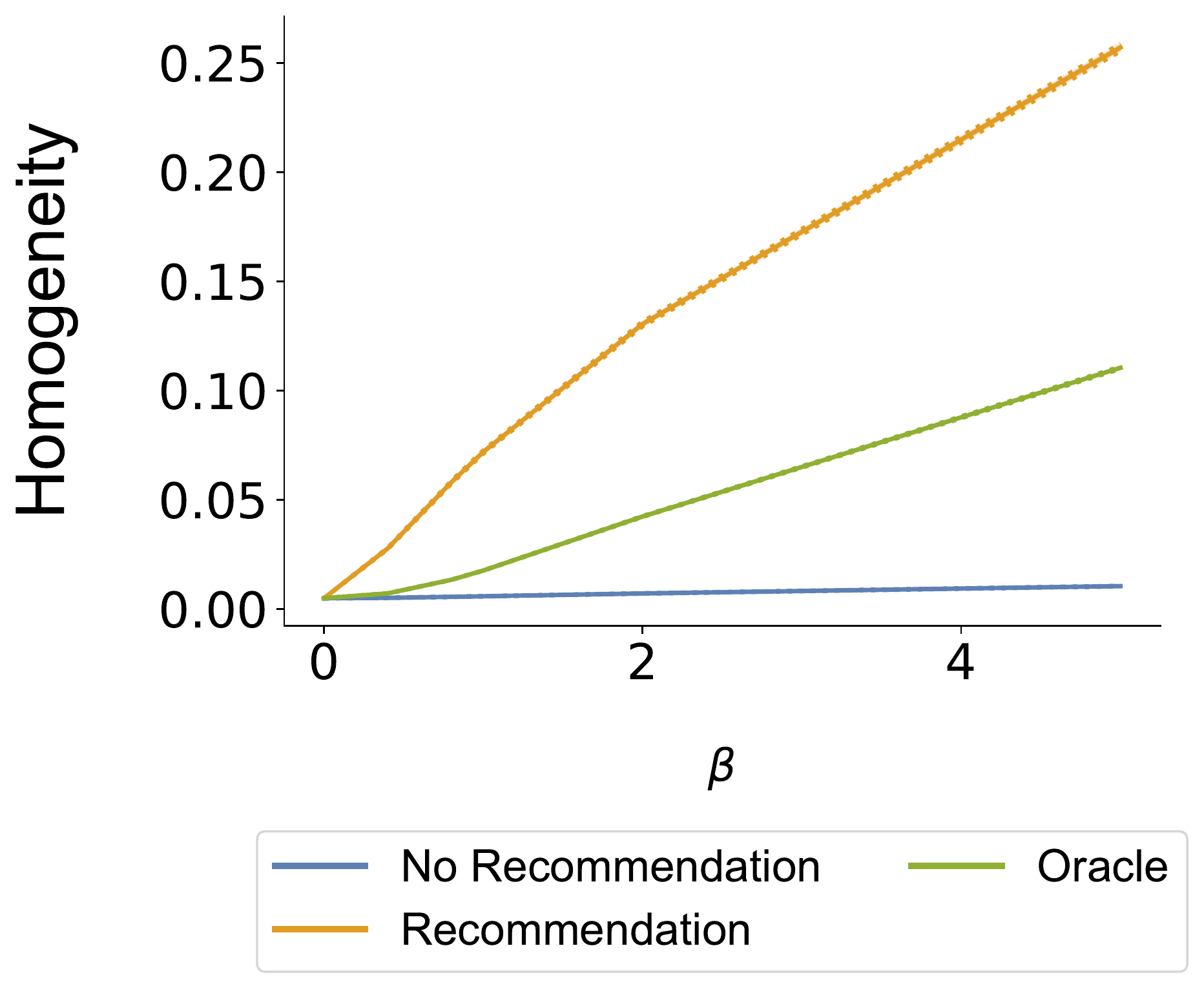} \hspace{1.0cm}
\includegraphics[width=.3\linewidth]{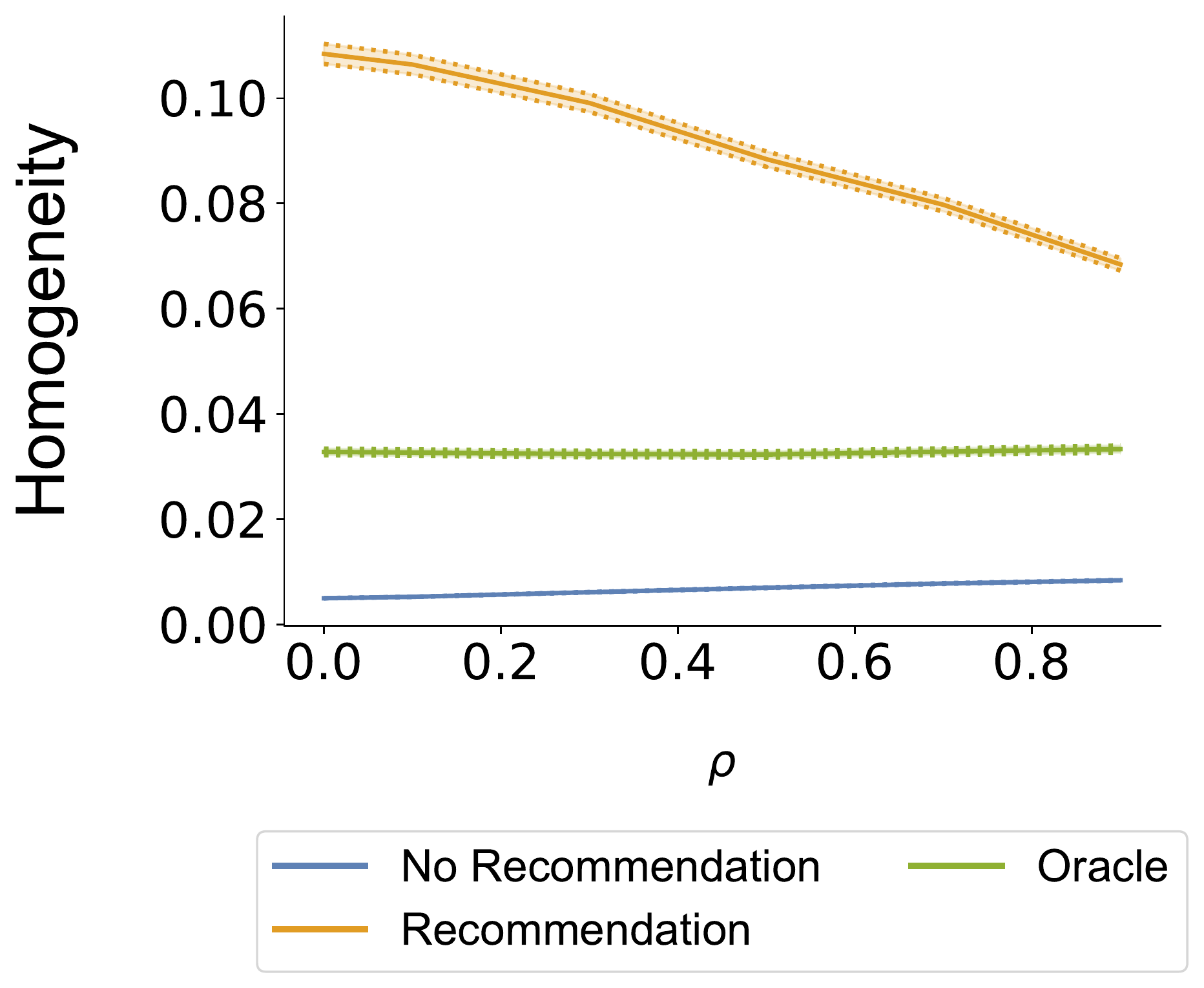}\label{fig:beta_homo}
\caption*{\scriptsize Notes: This figure displays the value of the homogeneity measure as we vary the weight of the common value component, $\beta$ (left) and correlation between valuations, $\rho$ (right). Each line represents this plot for a single recommendation regime. The shaded area represents the 95\% confidence interval.}
\end{figure}

First, we study how the degree of homogenization varies as we increase $\beta$, the weight of the common value component. As $\beta$ increases we expect that users should become increasingly homogeneous as the realized utilities of the items are now increasingly similar.~\autoref{fig:beta_homo} confirms that as the weight of the common-value component $\beta$ increases, users consume increasingly similar sets of items. The homogenization effect is strongest under the recommendation regime since the revelation of the common-value component induces users to consume items in similar areas of the product space. As $\beta$ increases, some amount of homogenization is optimal as can be seen from the oracle case. However, since users in the no-recommendation regime do not know the common-value component they engage in local consumption in different areas of the product space which leads to less than optimal homogeneity.
\par

We next study how the degree of homogeneity varies with $\rho$. ~\autoref{fig:beta_homo} shows how homogeneity decreases as $\rho$ increases in the recommendation regime. As was highlighted in Findings \ref{finding_local_consumption} and \ref{finding_diversity}, the degree of local consumption increases with $\rho$. Even though the revelation of the common-value component induces them to search in similar parts of the product space, their idiosyncratic components induce them to consume items in a more localized area of the product space as $\rho$ increases which leads to a decline in homogeneity.

\section{Recommender System Evaluation}
In this section we discuss how the insights from our model of user decision-making can inform the evaluation and design of recommender systems. The classic approach to evaluation is to predict user ratings for items and to compare how accurate this prediction is to recorded ratings data, either explicitly given by users or inferred from behavioral data. The RS should then recommend the items with the highest predicted ratings \cite{adomavicius2005toward}.
\par
There has been a recent movement away from such evaluation measures due to the observation that accurate recommendations are not necessarily useful recommendations \cite{mcnee2006being}. Our model illustrates a mechanism behind this observation. Consider the domain of movie recommendation and suppose a user has just watched the movie \textit{John Wick} and rated it highly. A RS attempting to predict user ratings may then predict that this user is very likely to enjoy the sequel, \textit{John Wick: Chapter Two}, as well. However, the user herself may also have made this inference since the two movies are very similar to each other. Thus, recommending this movie would not be not useful since the recommendation gives the user little information that she did not already know. The key insight is it is not useful since \textit{it ignores the inference the user themselves made and their updated beliefs}. The user may watch \textit{John Wick: Chapter Two}, then, even without recommendation, and the value of the recommendation was small.
\par
This intuition implies that RS should collect additional data beyond that which is traditionally recorded. The first and most crucial type of data to collect is individual user \textit{beliefs} about items that they have not yet consumed. As illustrated by our model, these beliefs are what drive the future consumption decisions of users and understanding these beliefs is crucial for determining the value of recommending certain items.\footnote{Additionally, user beliefs contain information that may not have been observed by the recommender that only observes user choices on the platform.} The second type of data that is relevant for RS designers to collect is how user beliefs change over time and, in particular, not just how individuals value the item they just consumed, but also how it impacts their beliefs about the quality of similar items.\footnote{Characterizing the similarity between items has been an important goal of designing content-based recommendations, though as noted by \cite{winecoff2019users}, how users perceive similarity between items is not always in line with how similarity is computed in content-based RS. Understanding how this impacts which items users update their beliefs about is an important direction for future work.} The third type of data is the risk-aversion levels of users as our model illustrates that the risk preferences of users are important for understanding what information RS can provide that materially leads users to alter their consumption patterns.
\par 
A natural follow-up question is how this additional data should be utilized in the design of good recommendations. Our model posits that recommendation provides value to users by providing them with information about the true valuation of an item if they were to consume it. Thus, the prediction problem for the recommender becomes predicting what item the user would choose with no recommendation and, correspondingly, what would be the most useful information to provide to the user that would lead her to consume a \textit{better} item than she would without recommendation. This links back to the intuition our model provided for the \textit{John Wick} example whereby collecting user beliefs and measuring how the user updated beliefs about similar items would lead the recommender to understand that the user would consume \textit{John Wick: Chapter Two}. Our approach would therefore imply that, with this as a starting point, the recommender's problem would be to predict what is the most useful information to give the user leading them to change the item that they eventually consume.
\par 
There have been a number of alternative recommendation evaluation metrics proposed in the literature with the aim of providing more useful recommendations than those provided by accuracy metrics, such as serendipity \cite{kotkov2016survey}, calibration \cite{steck2018calibrated}, coverage \cite{ge2010beyond}, novelty \cite{vargas2011rank}, and many others. Our approach most closely follows the set of proposed serendipity measures which are surveyed in \cite{kotkov2016survey}. As discussed by \cite{maksai2015predicting}, serendipitous recommendations are said to ``have the quality of being both unexpected and useful'' which is in line with the primary intuition behind our approach. The primary difference between our proposed approach and those existing in the literature is that ours crucially hinges on understanding user beliefs and the risk-preferences of users. For instance, \cite{vargas2011rank, kaminskas2014measuring} propose unexpectedness metrics that look at the dissimilarity of the proposed recommended items compared to what the recommender already knows the user likes. This metric depends only on the proposed item-set and not necessarily on the user's beliefs or how such a recommendation will change the item that the user consumes. \cite{kotkov2018investigating} provide a comprehensive overview of possible definitions of serendipity and ours is closest to their ``motivational novelty" definition, which is that the user was persuaded to consume an item as a result of recommendation.
\par 
Indeed, our approach allows us to give precise definitions for what it means for a recommendation to be \textit{unexpected} and \textit{useful} in the spirit of serendipitous recommendations. Our evaluation measure leads to useful recommendations since it leads users towards better items than they would consume without recommendation. It further results in ``unexpected'' recommendations since it explicitly incorporates user beliefs and thus allows the RS to understand how ``unexpected'' a recommendation would be from the perspective of a user. Finally, such a measure may lead to a perceived broadening of user preferences as has been discussed in \cite{herlocker2004evaluating, zhang2012auralist}. However, under our interpretation, it may be that their underlying preferences are unchanged and, instead, that recommendation and consumption themselves provides information that encourages users to explore different portions of the product space.
\newpage
\bibliographystyle{ACM-Reference-Format}
\bibliography{refs}

\end{document}